\documentclass[aps,pre,amsmath,amssymb,reprint,superscriptaddress]{revtex4-1}

\usepackage[utf8]{inputenc}
\usepackage{graphicx,sidecap}
\usepackage{subcaption}
\usepackage{booktabs}
\usepackage{hyperref}
\usepackage{tikz}

\hypersetup{colorlinks}

\newcommand{\vc}[1]{\mathbf{ #1}}
\newcommand{\p}[1]{\partial{#1}}
\newcommand{\rc}{r_{\mathrm{c}}}
\newcommand{\rd}{r_{\mathrm{d}}}
\renewcommand{\wr}{w_{\mathrm{\rho}}}

\newcommand{\Nw}{N_{\mathrm{w}}}

\newcommand{\bra}{\bar\rho^{(\alpha)}}
\newcommand{\brb}{\bar\rho^{(\beta)}}
\newcommand{\Baa}{B^{(\alpha\alpha)}}
\renewcommand{\Bbb}{B^{(\beta\beta)}}
\newcommand{\Bab}{B^{(\alpha\beta)}}

\begin{document}

\author{Peter Vanya}
\email{peter.vanya@gmail.com}
\affiliation{Department of Materials Science \& Metallurgy, University of Cambridge, 27 Charles Babbage Road, Cambridge CB3 0FS, United Kingdom}
\affiliation{Value for Money Unit, Ministry of Finance of the Slovak Republic, Štefanovičova 5, 817 82 Bratislava, Slovakia}

\author{James A. Elliott}
\email{jae1001@cam.ac.uk}
\affiliation{Department of Materials Science \& Metallurgy, University of Cambridge, 27 Charles Babbage Road, Cambridge CB3 0FS, United Kingdom}

\title{Definitions of local density in density-dependent potentials for mixtures}

\begin{abstract}
Density-dependent potentials are frequently used in materials simulations due to their approximate description of many-body effects at minimal computational cost. However, in order to apply such models to multi-component systems, an appropriate definition of total local particle density is required. Here, we discuss two definitions of local density in the context of many-body dissipative particle dynamics. We show that only a potential which combines local densities from all particle types in its argument gives physically meaningful results for all composition ratios. Drawing on the ideas from metal potentials, we redefine local density such that it can accommodate different inter-type interactions despite the constraint to keep the main interaction parameter constant, known as Warren's no-go theorem, and generalise the many-body potential to heterogeneous systems. 
We then show via simulation how liquid-liquid and liquid-solid coexistence can arise just by tuning the interaction parameters.
\end{abstract}

\maketitle

\section{Introduction}
Coarse-graining is a widely used approach across all scales to eliminate fast or unimportant degrees of freedom, speed up simulations and understand the most relevant physics. In electronic structure theory, core electrons are often coarse-grained into pseudopotentials; in classical atomistic simulations, all electrons are reduced to effective potentials between atoms; in soft matter, whole atoms and molecules can be collapsed into particles. These ``blobs" interact via effective pair potentials, which differ from classical, all-atom pair potentials in that they are parametrised only for a specific thermodynamic state. Thus, these potentials do not necessarily reproduce material properties at temperatures or pressures other than the one for which they were defined, which is commonly known as the transferability problem. 

A dependence on the local density (LD) of particles can be added to increase predictive abilities, motivated by the observation that some material properties cannot be captured purely by pair potentials~\cite{Louis_JPCM_2002, Merabia_JCP_2007}. In the solid state, LD is used to improve the description of metals and alloys. Coarse-graining electrons into an effective local electronic density significantly improves the description of fracture and the role of impurities. Examples of such metal potentials are the embedded atom method~\cite{Daw_PRL_1983, Daw_PRB_1984}, Finnis-Sinclair model~\cite{Finnis_PMA_1984} or Sutton-Chen potential~\cite{Sutton_PML_1990}. 

The general form of the energy in these metal potentials is:
\begin{equation}
U = \sum_{i<j} V(r_{ij}) + \sum_i u(\bar\rho_i),
\end{equation}
where $u(\bar\rho_i)$ is the self-energy of $i$th atom embedded in a local density $\bar\rho_i$ and $r_{ij} = |\vc r_i - \vc r_j|$ is the distance between the $i$th and $j$th atom at positions $\vc r_i$ and $\vc r_j$, respectively. The local density accounts for neighbouring particles via weight functions $\wr$ which decrease with distance:
\begin{equation}
\bar\rho_i = \sum_{j\neq i} \wr(r_{ij}).
\label{eq:loc_rho}
\end{equation}
For the Finnis-Sinclair and Sutton-Chen potentials, $u(\bar\rho_i) = -A\sqrt{\bar\rho_i}$, where the square root is motivated by the tight binding approximation. The force on $i$th atom is:
\begin{equation}
\vc F_i = -\frac{\p U}{\p{\vc r_i}} = - \sum_j \frac{\p U}{\p r_{ij}} \hat{\vc r}_{ij},
\label{eq:force_mdpd}
\end{equation}
where $\hat{\vc r}_{ij} = (\vc r_i - \vc r_j)/r_{ij}$ is a unit vector.

In soft matter, potentials with LD terms have been used either with a form defined \emph{a priori} in many-body dissipative particle dynamics (MDPD), or coarse-grained from the bottom-up, as with an application to mixing of water and benzene~\cite{Sanyal_JCP_2016, Sanyal_JPCB_2018}. Building on standard dissipative particle dynamics (DPD)~\cite{Espanol_JCP_2017}, MDPD is suitable for describing mesoscale systems with heterogeneous densities. This force field was introduced by Pagonabarraga and Frenkel~\cite{Pagonabarraga_JCP_2001} in general terms as a potential for non-ideal fluids and further developed by Trofimov \emph{et al.}~\cite{Trofimov_JCP_2002} and Warren~\cite{Warren_PRL_2001, Warren_PRE_2003}. MDPD was recently parametrised by the present authors for solvent mixtures~\cite{Vanya_PRE_2018}.

In the past decades, a significant body of research has been generated by applying standard or many-body DPD. The Karniadakis group investigated a range of applications including blood~\cite{Chang_PLOS_2016, Li_PTRSA_2014, Peng_PNAS_2013}, block copolymers~\cite{Li_MA_2009} and membranes~\cite{Li_Nanoscale_2012}. Ghoufi and Malfreyt explored extensively the vapour-liquid coexistence in MDPD~\cite{Ghoufi_PRE_2010, Ghoufi_PRE_2011, Ghoufi_EPJ_2013}. Merabia~\emph{et al.} investigated wetting of liquid on a solid substrate with a model similar to the present definition of MDPD~\cite{Merabia_EPJE_2006, Merabia_JNNFM_2008, Merabia_PRL_2008}. Another, related branch of research is smoothed-particle hydrodynamics accounting for thermal fluctuations and transport by discretising the Navier-Stokes equations~\cite{Espanol_PRE_2003, Thieulot_PRE_2005a, Thieulot_PRE_2005b, Litvinov_PRE_2008}.

Standard DPD has a purely repulsive pair potential with a cutoff yielding a force between two coarse-grained particles of the form:
\begin{equation}
\vc F_{ij} = A w(r_{ij}) \hat {\vc r}_{ij},
\end{equation}
where $A$ is an interaction parameter and $w(r)$ a weight function with a linear taper: $w(r) = 1-r/\rc$ for $r<\rc$ and 0 elsewhere. Together with the many-body term with self-energy $u(\bar\rho)$, it can be shown via eq.~\eqref{eq:force_mdpd} that the force from the MDPD potential is:
\begin{equation}
\vc F_{ij} = A w(r_{ij}) \hat{\vc r}_{ij} - [u'(\bar\rho_i) + u'(\bar\rho_j)] \wr'(r_{ij}) \hat{\vc r}_{ij}.
\end{equation}
This weight function has a cutoff and is normalised, such that $\int 4\pi r^2 \wr(r) dr = 1$.

The simplest form of self-energy is $u(r_i)=B\bar\rho_i^2/2$, with an interaction parameter $B$ and $\wr(r)\sim (1-r/\rd)^2$ for $r<\rd$ and 0 elsewhere~\footnote{In general, any power of the local density can be considered and thus the equation of state can be influenced, as was demonstrated by Trofimov \emph{et al.}~\cite{Trofimov_JCP_2002}.}. Setting $A<0$, $B>0$ and the many-body cutoff $\rd<\rc$ results in a potential that can produce a liquid-vapour coexistence, which makes MDPD applicable to systems containing interfaces between different phases.

\subsection{Multicomponent systems}
The generalisation of self-energy in MDPD to multicomponent systems has so far been ambiguous; the form of the LD proposed in the literature has been assumed without justification of the reasoning. For the $i$th particle, single-component LDs can be defined separately by particle type $\alpha$:
\begin{equation}
\bar\rho_i^{(\alpha)} = \sum_{j,j\in\alpha} \wr(r_{ij}).
\label{eq:loc_rho_2}
\end{equation}
So, for $n$-component systems there are $n$ different LDs for each particle. The ambiguity lies in the fact that it is not \emph{a priori} clear how to combine these correctly in the self-energy.

Mathematically, for pair potentials the energy per particle is a straightforward function of the coordinates of the neighbouring particles: $u(\{\vc r_i\})$. LD potentials have a function representing the LD and an outside wrapping function $u$: $u[\bar\rho(\{\vc r_i\})]$. In case of multicomponent systems, such as liquid mixtures or alloys, a question arises how to combine the terms within the wrapping function $u(x)$. Taking the simplest, two-component system composed of types $\alpha$ and $\beta$, two options are immediately apparent: (i.) $u[ \bar\rho^{(\alpha)}( \{\vc r_i\} ) + \bar\rho^{(\beta)}( \{\vc r_i\} ) ]$, and (ii.) $u[ \bar\rho^{(\alpha)}( \{\vc r_i\} )] + u[ \bar\rho^{(\beta)}( \{\vc r_i\} ) ]$~\footnote{There is also a trivial function where the self-energy contains only the local density of the particles of the like type, $u[\bar\rho^{(\alpha)}(\vc r_i)]$. However, the resulting force is the same as for the partial local density variant.}. We denote these as the partial and full LD variants respectively.

The partial LD variant was used by Sanyal \emph{et al.}~\cite{Sanyal_JPCB_2018}
and also had originally been implemented in the DL\_MESO package~\cite{DLMS}, which has inspired the present exploration. The total LD variant was discussed in Section V in Trofimov \emph{et al.}~\cite{Trofimov_JCP_2002} and, implicitly, in Warren~\cite{Warren_PRE_2013}, when introducing the no-go theorem stating that the parameter $B$ must be constant across particle types if the potential $U$ is to be conservative.

In this work, we discuss the viability of these two LD variants of multicomponent systems. We show that the partial variant can have type-dependent parameters $B^{(\alpha\beta)}$, but it behaves unphysically in that a simple relabelling of particles alters the forces between them. This in turn affects local ordering and phase behaviour. As a result, only the total LD variant is usable in practice. Drawing on research in solid state physics, we redefine the latter such that it can accommodate variability among different particle types. 

\begin{figure}
\centering
\includegraphics[width=0.4\textwidth]{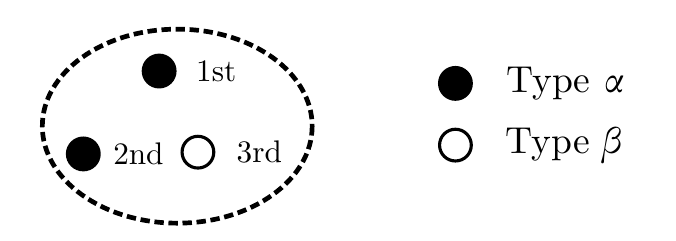}
\caption{A minimal multicomponent system to explore local density potentials.}
\label{fig:sys}
\end{figure}

To illustrate these points, we take a minimal two-component mixture of three particles and explicitly compute the forces between them for each of the variants of local density (Fig.~\ref{fig:sys}). We work with a general form of self-energy $u(\bar\rho_i)$ and weight function $\wr(r)$ and assume that all the particles are within the cutoff distance of one another. This reasoning also applies to other density-dependent potentials, not just MDPD. 

To proceed with algebra, we first define the necessary notation. In the most general case, there are three different interaction parameters in a two-component system. Like particles of type $\alpha$ and $\beta$ interact via parameters $B^{(\alpha\alpha)}$ and $B^{(\beta\beta)}$, respectively, and unlike particles via $B^{(\alpha\beta)}$. To treat the interaction parameter $B$ as an explicit prefactor, as in the case of metal potentials, we introduce a wrapping function $\psi$ such that $u(\bar\rho_i) = B\psi(\bar\rho_i)$. Commonly, $\psi(\bar\rho)$ is a polynomial, $\psi=\bar\rho^n/n$, with $n=2$ in MDPD.

\begin{table*}
\centering
\begin{ruledtabular}
\begin{tabular}{l l l l }
\toprule\toprule
Particle & Partial, type $\alpha$ & Partial, type $\beta$ & Total \\\hline
1 & $\bra_1 = \wr(r_{12})$ & $\brb_1 = \wr(r_{13})$ &
$\bar\rho_1 = \wr(r_{12}) + \wr(r_{13})$\\
2 & $\bra_2 = \wr(r_{12})$ & $\brb_2 = \wr(r_{23})$ &
$\bar\rho_2 = \wr(r_{12}) + \wr(r_{23})$\\
3 & $\bra_3 = \wr(r_{13}) + \wr(r_{23})$ & $\brb_3 = 0$ &
$\bar\rho_3 = \wr(r_{13}) + \wr(r_{23})$\\
\bottomrule\bottomrule
\end{tabular}
\end{ruledtabular}
\caption{List of partial local densities for each particle.}
\label{tbl:loc_rho}
\end{table*}

\section{Derivation of forces for partial local densities}
\label{sec:loc}
Starting with the partial LD variant, the local densities can be listed explicitly (Table~\ref{tbl:loc_rho}) for each of the three particles in the minimal system. From these, the self-energies of the particles follow:
\begin{align}
u_1 &= \Baa \psi(\bra_1) + \Bab \psi(\brb_1),\\
u_2 &= \Baa \psi(\bra_2)+ \Bab \psi(\brb_2),\\
u_3 &= \Bab \psi(\bra_3) + \Bbb \psi(\brb_3).
\end{align}
The total energy $U=u_1+u_2+u_3$. Computing, e.g., $\p U/\p{\vc r_1}$:
\begin{multline}
\frac{\p U}{\p{\vc r_1}} = 
\Baa [\psi'(\bra_1) + \psi'(\bra_2)] \wr'(r_{12}) \hat{\vc r}_{12} \\
+ \Bab [\psi'(\brb_1) + \psi'(\bra_3)] \wr'(r_{13}) \hat{\vc r}_{13},
\label{dudr1}
\end{multline}
$\p U/\p{\vc r_2}$ can be obtained from eq.~\eqref{dudr1} by simply transposing particle indices 1 and 2 and:
\begin{multline}
\frac{\p U}{\p{\vc r_3}} = 
\Bab [\psi(\brb_1) + \psi'(\bra_3)] \wr'(r_{12}) \hat{\vc r}_{31} \\
+ \Bab [\psi'(\brb_2) + \psi(\bra_3)] \wr'(r_{23}) \hat{\vc r}_{32}.
\end{multline}

Every force $\vc F_{ij}$ has the form of eq.~\eqref{eq:force_mdpd} and $\vc F_{ij} = - \vc F_{ji}$ for every pair $\{i,j\}$ in line with Newton's third law. Hence, the self-energy of the partial LD variant is conservative and, at the same time, allows for type-specific interaction parameters $B$.

\section{Problem with particle relabelling}
The freedom to use different parameters for unlike types provided by the partial LD variant can be important for an appropriate depiction of phase behaviour of mixtures. However, a new problem arises: the interaction strength of particles of unlike types is artificially lowered only due to the fact that they have different labels, not due to physical differences.

In a homogeneous, single-component MDPD liquid with parameter $B$, the local density is same for every particle and equal to the global density, $\bar\rho_i \approx \rho$ (assuming mean-field approximation). The force between any two particles is then:
\begin{align}
\vc F_{ij} & \approx B[\psi'(\rho) + \psi'(\rho)]\wr'(r_{ij}) \hat{\vc r}_{ij} \nonumber \\
& = 2B\psi'(\rho)\wr'(r_{ij})\hat{\vc r}_{ij}.
\end{align}

Consider randomly splitting all the particles into two types but keeping the interaction parameter constant, $\Baa=\Bab=\Bbb=B$. Now, every particle sees around itself, on average, one half of the particles of type $\alpha$ and the other half of type $\beta$, as the system remains physically the same and hence perfectly mixed. Therefore, the average local density of both type $\alpha$ and $\beta$ particles is $\bar\rho_i^{(\alpha/\beta)}\approx \rho/2$. Computing the force between like type particles $i$ and $j$ yields:
\begin{align}
\vc F_{ij} &= B[\psi'(\rho/2) + \psi'(\rho/2)]\wr'(r_{ij}) \hat{\vc r}_{ij} \nonumber\\
&= 2B\psi'(\rho/2)\wr'(r_{ij})\hat{\vc r}_{ij}.
\end{align}
These forces are \emph{not} equal, since generally $\psi'(\rho)\neq\psi'(\rho/2)$. The only exception is the case when $\psi$ depends linearly on $\bar\rho$ and $\psi'\sim 1$, which is the force field of standard DPD. With the simplest non-trivial definition of self-energy, $\psi(\bar\rho)=\bar\rho^2/2$, the force on any particle would become twice as small purely due to relabelling, and, in simulations of mixtures with $m\gg 1$ components, $m$-times smaller.

\begin{figure}
\centering
\includegraphics[width=0.45\textwidth]{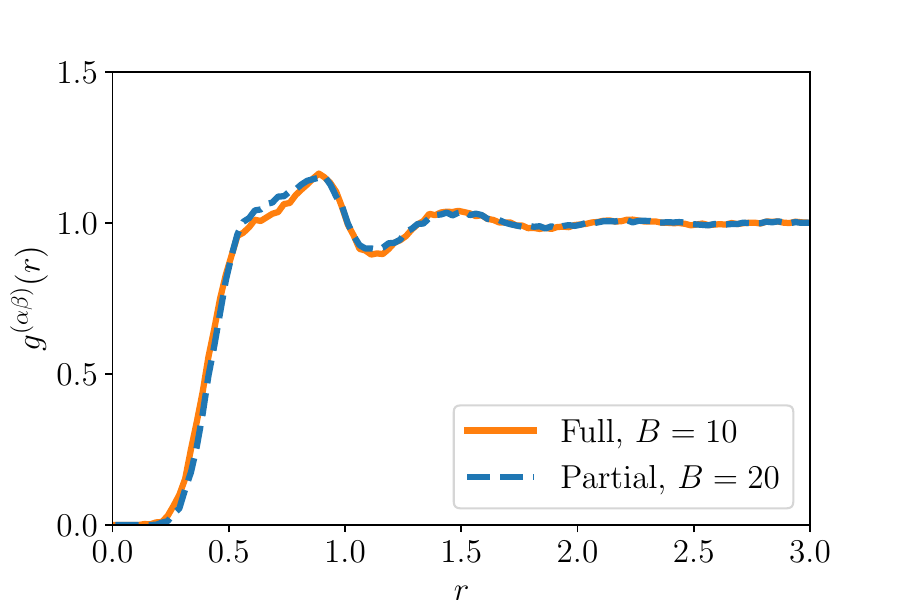}
\caption{Inter-type radial distribution functions of two variants of local density demonstrating the unphysical behaviour of the partial variant. The slight deviation is most likely due to the different temperatures at which these systems equilibrate.}
\label{fig:rdf}
\end{figure}

This flaw is manifest in the local structure of a liquid, which is represented by the radial distribution function (RDF). Using DL\_MESO version 2.6~\cite{DLMS}, we set up two simulations of an effectively single-component liquid with arbitrarily relabelled particles, exploring the full LD variant with parameter $B=10$ and partial LD variant with $B=20$, which should be different liquids. In each case we measured $g^{(\alpha\beta)}(r)$. Fig.~\ref{fig:rdf} shows the near identity of these two RDFs, demonstrating that the partial LD variant artificially lowers the interaction by a factor of two.

\begin{figure}[h]
\centering
\begin{subfigure}[b]{0.24\textwidth}
\includegraphics[width=1\textwidth]{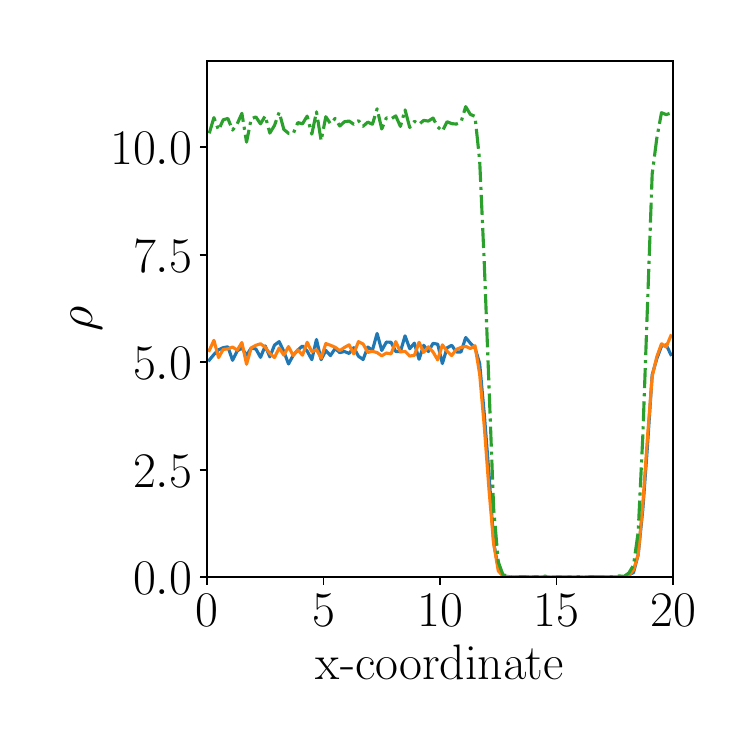}
\caption{}
\label{fig:rho2}
\end{subfigure}
\hspace{-4mm}
\begin{subfigure}[b]{0.24\textwidth}
\includegraphics[width=1\textwidth]{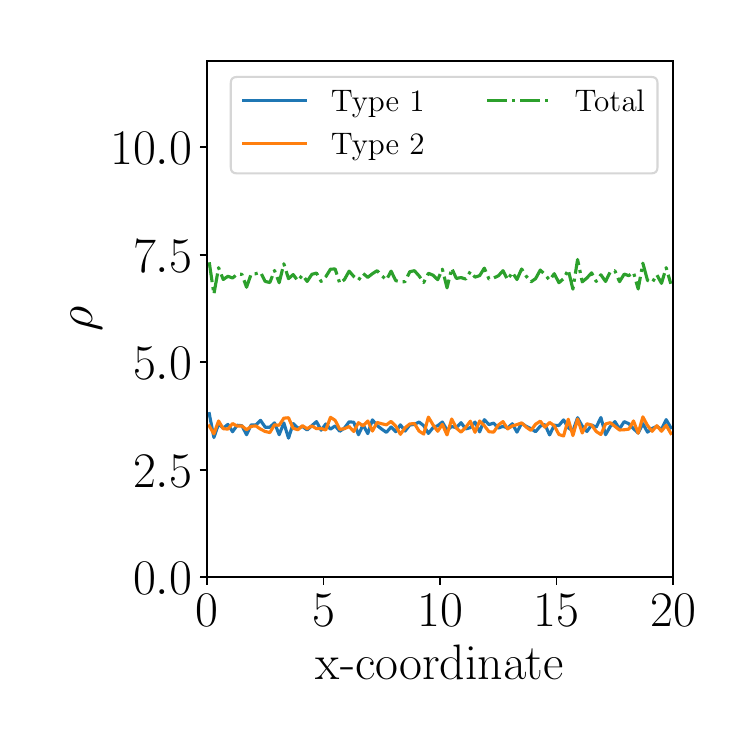}
\caption{}
\label{fig:rho1}
\end{subfigure}
\caption{Density profiles of a homogeneous liquid with particles randomly split into two types, for (a) partial and (b) total local density variant. The former produces an unphysically high global density and a spurious vapour phase.  The values of $\rho$ and $x$ are given in DPD units.}
\label{fig:rho}
\end{figure}

To illustrate the practical consequences of this, we consider a binary liquid interacting via an MDPD potential where interaction parameters are constant across particle types, $A=-18$ and $B=8$. These values represent water at a coarse-graining degree 6 with equilibrium density $\rho_0=6.70$~\cite{Vanya_JCP_2019}. Setting the density to $\rho=7>\rho_0$, this effectively single-component liquid with arbitrarily relabelled particles should homogeneously fill the simulation cell. We simulated the two local density variants and composed density profiles for each type after equilibration. Fig.~\ref{fig:rho} shows that the expected phase behaviour for a homogeneous liquid is only reproduced with the total LD variant.

\section{Generalised many-body force field}
\subsection{Derivation of forces for total local densities}
Having shown that the partial LD variant produces unphysical behaviour, only the total LD variant, which works with $\bar\rho_i =\bra_i + \brb_i$, remains. However, Warren's no-go theorem states that, for a type-independent definition of local density (eq.~\eqref{eq:loc_rho_2}), only a type-independent parameter $B$ is allowed, \emph{i.e.} $u_i=B\psi(\bar\rho_i)$ $\forall i$~\cite{Warren_PRE_2013}. This means that the self-energy for the force on, e.g., 1st particle with the following form:
\begin{multline}
\vc F_1 = -\Baa (\psi'(\bar\rho_1) + \psi'(\bar\rho_2)) \wr'(r_{12}) \hat{\vc r}_{12} \\
-\Bab (\psi'(\bar\rho_1) + \psi'(\bar\rho_3)) \wr'(r_{13}) \hat{\vc r}_{13}.
\end{multline}
does \emph{not} exist unless $\Baa=\Bab=\Bbb$. This constraint on $B$ means that it is not possible to distinguish particles in MDPD purely by the many-body potential term, which limits the versatility of this method.

Borrowing from the formalism of metal potentials, we show that there is a way to preserve type-dependent forces by introducing a type-dependent local density~\cite{RafiiTabar_PML_1991,Johnson_PRB_1989}. Generally the local density for particle of type $\alpha$ accounting for neighbouring particles of type $\beta$ is:
\begin{equation}
\bar\rho^{(\alpha\beta)}_i = \sum_{j} \wr^{(\alpha\beta)} (r_{ij}),
\quad i\in\alpha,
\quad j\in\beta.
\end{equation}

For a two-component system, there are four possible local densities: $\bar\rho^{(\alpha\alpha)}, \bar\rho^{(\alpha\beta)}, \bar\rho^{(\beta\alpha)},\bar\rho^{(\beta\beta)}$. Generally, the influence of a particle of type $\alpha$ to the local density of a particle of type  $\beta$ might not be the same, but in practice $\bar\rho^{(\alpha\beta)}= \bar\rho^{(\beta\alpha)}$. 

Considering again the minimal three-particle system (Fig.~\ref{fig:sys}), the self-energies are:
\begin{align}
u_1 &= B \psi \left(\bar\rho_1^{(\alpha\alpha)} + \bar\rho_1^{(\alpha\beta)} \right) = B\psi(\bar\rho_1),\\
u_2 &= B \psi \left(\bar\rho_2^{(\alpha\alpha)} + \bar\rho_2^{(\alpha\beta)} \right) = B\psi(\bar\rho_2),\\
u_3 &= B \psi \left(\bar\rho_3^{(\alpha\beta)} + \bar\rho_3^{(\beta\beta)} \right) = B\psi(\bar\rho_3),
\end{align}
and the total energy $U=u_1+u_2+u_3$. Differentiating to obtain the force on particle 1 yields:
\begin{multline}
\vc F_1 = 
- B [\psi'(\bar\rho_1) + \psi'(\bar\rho_2)] \wr'^{(\alpha\alpha)} (r_{12}) \hat{\vc r}_{12} \\
- B [\psi'(\bar\rho_1) + \psi'(\bar\rho_3)] \wr'^{(\alpha\beta)} (r_{13}) \hat{\vc r}_{13} \\
= -\vc F_{12} - \vc F_{13},
\end{multline}
and the forces on forces on particles 2 and 3 can be obtained similarly.

The cross-type interaction is now represented by the term $B\wr'^{(\alpha\beta)}$, which gives the freedom to tune it via the derivative of the weight function $\wr^{(\alpha\beta)}$. A simple arithmetic mixing rule can be used: $\wr^{(\alpha\beta)}= \left(\wr^{(\alpha\alpha)} + \wr^{(\beta\beta)} \right)/2$, but a more general form might be required for coarse-grained systems, possibly involving the Flory-Huggins $\chi$-parameter as in the case of standard DPD. (A geometric mixing, as in the case of metal potentials~\cite{RafiiTabar_PML_1991}, might not be suitable due to the explicit and necessary cutoff of coarse-grained potentials.)

\subsection{Mixing of different material types}
Finally, we address the mixing of different materials. An example is a liquid or a polymer on metal surface, which is a typical setting for heterogeneous catalysis and so is of immense practical importance. 

The total LD self-energy allows for definitions of the wrapping function $\psi$ depending on the material phase. Referring to Fig.~\ref{fig:sys}, consider the first two particles metallic (M) and the third a liquid (L). Each phase has a different wrapping function: $\psi_{\rm M}$ for a metal can be a square root, and $\psi_{\rm L}$ for a liquid can be a square to allow for liquid-vapour coexistence. The self-energies are:
\begin{align}
u_1 &= B \psi_{\rm M} \left(\bar\rho_1^{(\alpha\alpha)} + \bar\rho_1^{(\alpha\beta)} \right) = 
B\psi_{\rm M} (\bar\rho_1),\\
u_2 &= B \psi_{\rm M} \left(\bar\rho_2^{(\alpha\alpha)} + \bar\rho_2^{(\alpha\beta)} \right) = 
B\psi_{\rm M} (\bar\rho_2),\\
u_3 &= B \psi_{\rm L} \left(\bar\rho_3^{(\alpha\beta)} + \bar\rho_3^{(\beta\beta)} \right) = 
B\psi_{\rm L} (\bar\rho_3).
\label{eq:hetero}
\end{align}
Following the previous section, the force on particle 1 is:
\begin{multline}
\vc F_1 = 
-B [\psi_{\rm M}'(\bar\rho_1) + \psi_{\rm M}'(\bar\rho_2)] \wr'^{(\alpha\alpha)} (r_{12}) \hat{\vc r}_{12} \\ - B [\psi_{\rm M}'(\bar\rho_1) + \psi_{\rm L}'(\bar\rho_3)] \wr'^{(\alpha\beta)} (r_{13}) \hat{\vc r}_{13},
\end{multline}
where $\vc F_{13}$ contains one term from neighbouring metal atoms, $\psi'_{\rm M}$, and another from neighbouring liquid particles, $\psi'_{\rm L}$. The cross-type weight function $\wr^{(\alpha\beta)}$ requires further investigation to determine its appropriate form or a range of forms, but at this stage it is sufficient to note that, as before, it represents unambiguously the material-dependent interaction via the term $B\wr'^{(\alpha\beta)}$.

\subsection{Exploring new force fields via simulation}
To demonstrate the versatility of this generalised MDPD force field for describing inhomogeneous mixtures, we perform several simulations using a custom-written code~\footnote{Available at \url{https://github.com/petervanya/DPDsim}.}.

\begin{figure}
\centering
\includegraphics[width=0.48\textwidth]{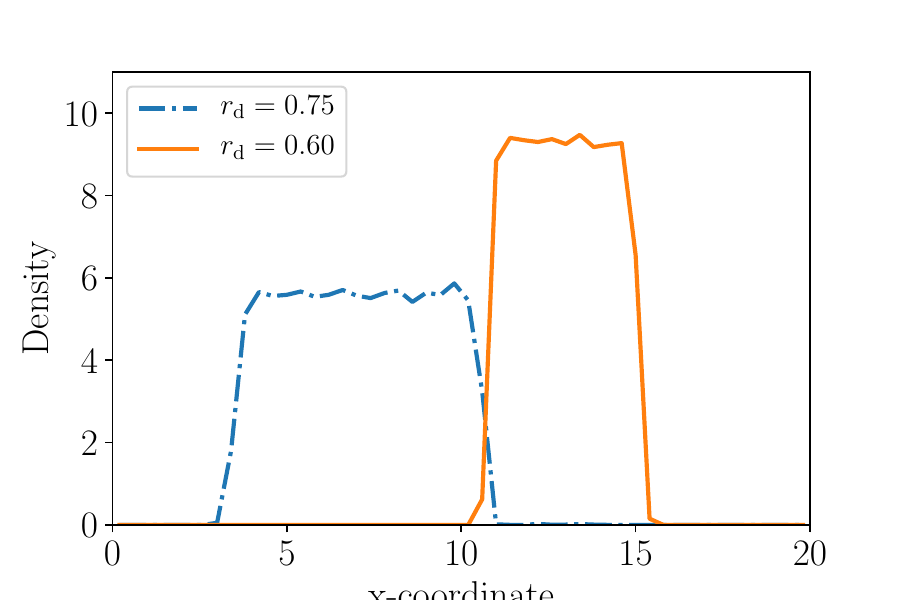}
\caption{A coexistence of inhomogeneous liquids using a generalised MDPD force field allowing for different many-body cutoffs.}
\label{fig:gmdpd_1}
\end{figure}

Before doing so, it is necessary to redefine the repulsion parameter $B$ to follow the logic of the embedded atom method using unique $B$'s but varying parameters within the wrapping functions with particle type. The convention with MDPD has been so far to define $B$ through the force: $F_{ij}=B_{\rm o} (\bar\rho_i + \bar\rho_j)(1-r/\rd)$ with $r<\rd$~\cite{Warren_PRE_2003}. Here, we start from the energy, $u_i = B \bar\rho_i^2 / 2$, so $B$ is modified by the normalisation factor of the weight function; the force for like-type particles then becomes $F_{ij} = 15/(\pi\rd^4) \, B\, (\bar\rho_i + \bar\rho_j)(1-r/\rd)$, so $B=B_{\rm o} \, \pi \rd^4/15$. For a typical value of $B_{\rm o}=30$, $B\approx 2$ for $\rd=0.75$.

Using this convention, we first investigate inhomogeneous liquids with same quadratic wrapping function for both particle types, $\psi(x) = x^2/2$, with parameters $A=-40, B=2$. The cross-type repulsion is represented by decreased attraction $A_{12} = A + \Delta A$, setting $\Delta A=20$. The difference between the liquids is marked by many-body cutoff set to 0.75 and 0.60 for type 1 and type 2, respectively, and cross-interaction as the arithmetic mean of these, 0.675. The simulation contained 2000 particles, split equally between the two types, in a $20 \times 5 \times 5$ cell and was run for 200 reduced time units with timestep $\Delta t=0.01$. Figure~\ref{fig:gmdpd_1} shows a coexistence of two separated phases at different densities, a feature so far unavailable in MDPD.

\begin{figure}[b]
\includegraphics[width=0.48\textwidth]{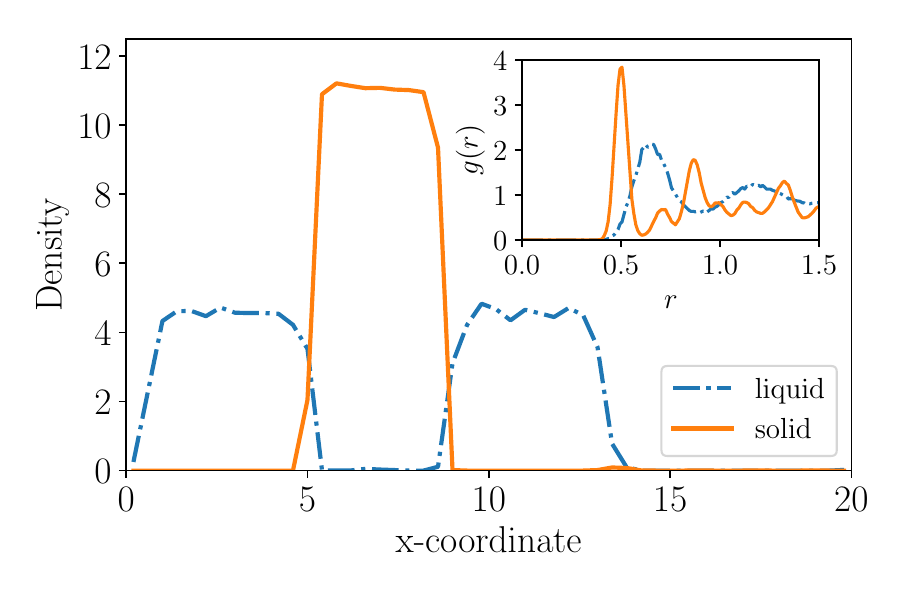}
\caption{A coexistence of liquid and solid phase using a generalised MDPD force field achieved by choosing different many-body cutoffs and wrapping functions. The inset shows radial distribution functions; the tall first peak of the orange curve is a signature of the solid phase.}
\label{fig:gmdpd_2}
\end{figure}

As a second simulation, following eq.~\eqref{eq:hetero} we consider a system with wrapping functions differing with particle types. For simplicity, we constrain the choice to polynomial functions: $\psi(x) = x^{\Nw}/\Nw$. We chose $\Nw=2$ for phase 1 and $\Nw=3$ for phase 2, the repulsion parameter $B=3$, and many-body cutoffs $\rd^{(11)} = 0.75$, $\rd^{(22)}=0.55$, and $\rd^{(12)}=0.65$. We simulated for the same time period as before using a smaller timestep, $\Delta t = 0.006$ in order to retain the temperature within 10\% of $k_{\rm B} T=1$. This setup shows the coexistence of a liquid and a solid phase (Fig.~\ref{fig:gmdpd_2}). The solid phase is identified by a radial distribution function with a sharp first peak and more frequent oscillations, in contrast to that for the liquid phase, which is flatter and smoother (in the inset of Fig.~\ref{fig:gmdpd_2}). This can be compared with Fig.~5a of our previous work~\cite{Vanya_PRE_2018}.

With these exploratory simulations, we note several observations. Firstly, is not advisable to use $\Nw=1/2$, as in the embedded atom method, due to the finiteness of the MDPD potential. This feature produces, with some non-negligible probability, vanishing local densities, from which, after taking the derivative to compute forces, results in a zero in the denominator. A solution could be to increase the range, but this would go against the spirit of coarse-graining and simulation efficiency. Secondly, with increasing $\Nw$ it is necessary to minimise the box prior to the simulation run; such a need does not arise with lower values due to fast equilibration. Thirdly, it is also possible to achieve a liquid-solid phase coexistence using purely $\Nw=2$ for both particle types and controlling the difference between the phases only by the many-body cutoff. Fourth, we only set the cross-type repulsion via the interaction parameter $A$ and many-body cutoff $\rd$; there are other options to modify the many-body part of the interaction apart from the parameter $\rd$ that could provide further flexibility to describe real materials; these remain to be investigated.

Finally, there is the question of applications of the generalised MDPD force field to real materials. Standard DPD as well as many-body DPD relied on \emph{ad hoc} parametrisation dependent on the choice of the material by setting the reduced length scale based on its molecular volume. However, each component of a heterogeneous mixture has its own scale, so a new parametrisation protocol must be devised. As the atomistic resolution, on which the EAM is based, is not available, and the metallic faces and lattice constants are ambiguous concepts in coarse-graining, an open question remains about what the default physical properties for the parametrisation of the solid state should be. It could be possible to use the total energy and compressibility.

\section{Conclusion}
The definition of local density for many-body, density-dependent potentials must include all particle types. Otherwise, a mathematical relabelling of particles yields unphysical results  unless the potential energy is redefined. Following the idea of alloys in metal potentials, we showed that the total local density composed of contributions from all particles can be distinguished by the particle type and is able to account for different cross-type interactions despite the need to keep the main interaction parameter constant to obey Warren's no-go theorem~\cite{Warren_PRE_2013}. 

We generalised the many-body coarse-grained potential and demonstrated its usefulness via simulations of a mixture of liquids with unequal densities and a liquid-solid system. This force field thus opens up avenues for treatment of material combinations such as soft matter on metal surface with potential applications in heterogeneous catalysis.

\section{Acknowledgments}
The authors thank Johnson Matthey and the Engineering and Physical Sciences Research Council (EPSRC) for financial support and So\v{n}a Slobodníková for inspiring scientific discussions.

\bibliography{ref.bib}

\begin{thebibliography}{37}%
\makeatletter
\providecommand \@ifxundefined [1]{%
 \@ifx{#1\undefined}
}%
\providecommand \@ifnum [1]{%
 \ifnum #1\expandafter \@firstoftwo
 \else \expandafter \@secondoftwo
 \fi
}%
\providecommand \@ifx [1]{%
 \ifx #1\expandafter \@firstoftwo
 \else \expandafter \@secondoftwo
 \fi
}%
\providecommand \natexlab [1]{#1}%
\providecommand \enquote  [1]{``#1''}%
\providecommand \bibnamefont  [1]{#1}%
\providecommand \bibfnamefont [1]{#1}%
\providecommand \citenamefont [1]{#1}%
\providecommand \href@noop [0]{\@secondoftwo}%
\providecommand \href [0]{\begingroup \@sanitize@url \@href}%
\providecommand \@href[1]{\@@startlink{#1}\@@href}%
\providecommand \@@href[1]{\endgroup#1\@@endlink}%
\providecommand \@sanitize@url [0]{\catcode `\\12\catcode `\$12\catcode
  `\&12\catcode `\#12\catcode `\^12\catcode `\_12\catcode `\%12\relax}%
\providecommand \@@startlink[1]{}%
\providecommand \@@endlink[0]{}%
\providecommand \url  [0]{\begingroup\@sanitize@url \@url }%
\providecommand \@url [1]{\endgroup\@href {#1}{\urlprefix }}%
\providecommand \urlprefix  [0]{URL }%
\providecommand \Eprint [0]{\href }%
\providecommand \doibase [0]{http://dx.doi.org/}%
\providecommand \selectlanguage [0]{\@gobble}%
\providecommand \bibinfo  [0]{\@secondoftwo}%
\providecommand \bibfield  [0]{\@secondoftwo}%
\providecommand \translation [1]{[#1]}%
\providecommand \BibitemOpen [0]{}%
\providecommand \bibitemStop [0]{}%
\providecommand \bibitemNoStop [0]{.\EOS\space}%
\providecommand \EOS [0]{\spacefactor3000\relax}%
\providecommand \BibitemShut  [1]{\csname bibitem#1\endcsname}%
\let\auto@bib@innerbib\@empty
\bibitem [{\citenamefont {Louis}(2002)}]{Louis_JPCM_2002}%
  \BibitemOpen
  \bibfield  {author} {\bibinfo {author} {\bibfnamefont {A.~A.}\ \bibnamefont
  {Louis}},\ }\href {\doibase 10.1088/0953-8984/14/40/311} {\bibfield
  {journal} {\bibinfo  {journal} {Journal of Physics: Condensed Matter}\
  }\textbf {\bibinfo {volume} {14}},\ \bibinfo {pages} {9187} (\bibinfo {year}
  {2002})}\BibitemShut {NoStop}%
\bibitem [{\citenamefont {Merabia}\ and\ \citenamefont
  {Pagonabarraga}(2007)}]{Merabia_JCP_2007}%
  \BibitemOpen
  \bibfield  {author} {\bibinfo {author} {\bibfnamefont {S.}~\bibnamefont
  {Merabia}}\ and\ \bibinfo {author} {\bibfnamefont {I.}~\bibnamefont
  {Pagonabarraga}},\ }\href {\doibase 10.1063/1.2751496} {\bibfield  {journal}
  {\bibinfo  {journal} {The Journal of Chemical Physics}\ }\textbf {\bibinfo
  {volume} {127}},\ \bibinfo {pages} {054903} (\bibinfo {year}
  {2007})}\BibitemShut {NoStop}%
\bibitem [{\citenamefont {Daw}\ and\ \citenamefont
  {Baskes}(1983)}]{Daw_PRL_1983}%
  \BibitemOpen
  \bibfield  {author} {\bibinfo {author} {\bibfnamefont {M.~S.}\ \bibnamefont
  {Daw}}\ and\ \bibinfo {author} {\bibfnamefont {M.~I.}\ \bibnamefont
  {Baskes}},\ }\href {\doibase 10.1103/PhysRevLett.50.1285} {\bibfield
  {journal} {\bibinfo  {journal} {Phys. Rev. Lett.}\ }\textbf {\bibinfo
  {volume} {50}},\ \bibinfo {pages} {1285} (\bibinfo {year}
  {1983})}\BibitemShut {NoStop}%
\bibitem [{\citenamefont {Daw}\ and\ \citenamefont
  {Baskes}(1984)}]{Daw_PRB_1984}%
  \BibitemOpen
  \bibfield  {author} {\bibinfo {author} {\bibfnamefont {M.~S.}\ \bibnamefont
  {Daw}}\ and\ \bibinfo {author} {\bibfnamefont {M.~I.}\ \bibnamefont
  {Baskes}},\ }\href {\doibase 10.1103/PhysRevB.29.6443} {\bibfield  {journal}
  {\bibinfo  {journal} {Phys. Rev. B}\ }\textbf {\bibinfo {volume} {29}},\
  \bibinfo {pages} {6443} (\bibinfo {year} {1984})}\BibitemShut {NoStop}%
\bibitem [{\citenamefont {Finnis}\ and\ \citenamefont
  {Sinclair}(1984)}]{Finnis_PMA_1984}%
  \BibitemOpen
  \bibfield  {author} {\bibinfo {author} {\bibfnamefont {M.~W.}\ \bibnamefont
  {Finnis}}\ and\ \bibinfo {author} {\bibfnamefont {J.~E.}\ \bibnamefont
  {Sinclair}},\ }\href {\doibase 10.1080/01418618408244210} {\bibfield
  {journal} {\bibinfo  {journal} {Philosophical Magazine A}\ }\textbf {\bibinfo
  {volume} {50}},\ \bibinfo {pages} {45} (\bibinfo {year} {1984})}\BibitemShut
  {NoStop}%
\bibitem [{\citenamefont {Sutton}\ and\ \citenamefont
  {Chen}(1990)}]{Sutton_PML_1990}%
  \BibitemOpen
  \bibfield  {author} {\bibinfo {author} {\bibfnamefont {A.~P.}\ \bibnamefont
  {Sutton}}\ and\ \bibinfo {author} {\bibfnamefont {J.}~\bibnamefont {Chen}},\
  }\href {\doibase 10.1080/09500839008206493} {\bibfield  {journal} {\bibinfo
  {journal} {Philosophical Magazine Letters}\ }\textbf {\bibinfo {volume}
  {61}},\ \bibinfo {pages} {139} (\bibinfo {year} {1990})}\BibitemShut
  {NoStop}%
\bibitem [{\citenamefont {Sanyal}\ and\ \citenamefont
  {Shell}(2016)}]{Sanyal_JCP_2016}%
  \BibitemOpen
  \bibfield  {author} {\bibinfo {author} {\bibfnamefont {T.}~\bibnamefont
  {Sanyal}}\ and\ \bibinfo {author} {\bibfnamefont {M.~S.}\ \bibnamefont
  {Shell}},\ }\href {\doibase 10.1063/1.4958629} {\bibfield  {journal}
  {\bibinfo  {journal} {The Journal of Chemical Physics}\ }\textbf {\bibinfo
  {volume} {145}},\ \bibinfo {pages} {034109} (\bibinfo {year}
  {2016})}\BibitemShut {NoStop}%
\bibitem [{\citenamefont {Sanyal}\ and\ \citenamefont
  {Shell}(2018)}]{Sanyal_JPCB_2018}%
  \BibitemOpen
  \bibfield  {author} {\bibinfo {author} {\bibfnamefont {T.}~\bibnamefont
  {Sanyal}}\ and\ \bibinfo {author} {\bibfnamefont {M.~S.}\ \bibnamefont
  {Shell}},\ }\href {\doibase 10.1021/acs.jpcb.7b12446} {\bibfield  {journal}
  {\bibinfo  {journal} {The Journal of Physical Chemistry B}\ }\textbf
  {\bibinfo {volume} {122}},\ \bibinfo {pages} {5678} (\bibinfo {year}
  {2018})},\ \bibinfo {note} {pMID: 29466859}\BibitemShut {NoStop}%
\bibitem [{\citenamefont {Espa{\~n}ol}\ and\ \citenamefont
  {Warren}(2017)}]{Espanol_JCP_2017}%
  \BibitemOpen
  \bibfield  {author} {\bibinfo {author} {\bibfnamefont {P.}~\bibnamefont
  {Espa{\~n}ol}}\ and\ \bibinfo {author} {\bibfnamefont {P.~B.}\ \bibnamefont
  {Warren}},\ }\href {\doibase 10.1063/1.4979514} {\bibfield  {journal}
  {\bibinfo  {journal} {The Journal of Chemical Physics}\ }\textbf {\bibinfo
  {volume} {146}},\ \bibinfo {pages} {150901} (\bibinfo {year}
  {2017})}\BibitemShut {NoStop}%
\bibitem [{\citenamefont {Pagonabarraga}\ and\ \citenamefont
  {Frenkel}(2001)}]{Pagonabarraga_JCP_2001}%
  \BibitemOpen
  \bibfield  {author} {\bibinfo {author} {\bibfnamefont {I.}~\bibnamefont
  {Pagonabarraga}}\ and\ \bibinfo {author} {\bibfnamefont {D.}~\bibnamefont
  {Frenkel}},\ }\href {\doibase 10.1063/1.1396848} {\bibfield  {journal}
  {\bibinfo  {journal} {The Journal of Chemical Physics}\ }\textbf {\bibinfo
  {volume} {115}},\ \bibinfo {pages} {5015} (\bibinfo {year}
  {2001})}\BibitemShut {NoStop}%
\bibitem [{\citenamefont {Trofimov}\ \emph {et~al.}(2002)\citenamefont
  {Trofimov}, \citenamefont {Nies},\ and\ \citenamefont
  {Michels}}]{Trofimov_JCP_2002}%
  \BibitemOpen
  \bibfield  {author} {\bibinfo {author} {\bibfnamefont {S.~Y.}\ \bibnamefont
  {Trofimov}}, \bibinfo {author} {\bibfnamefont {E.~L.~F.}\ \bibnamefont
  {Nies}}, \ and\ \bibinfo {author} {\bibfnamefont {M.~a.~J.}\ \bibnamefont
  {Michels}},\ }\href {\doibase 10.1063/1.1515774} {\bibfield  {journal}
  {\bibinfo  {journal} {The Journal of Chemical Physics}\ }\textbf {\bibinfo
  {volume} {117}},\ \bibinfo {pages} {9383} (\bibinfo {year}
  {2002})}\BibitemShut {NoStop}%
\bibitem [{\citenamefont {Warren}(2001)}]{Warren_PRL_2001}%
  \BibitemOpen
  \bibfield  {author} {\bibinfo {author} {\bibfnamefont {P.~B.}\ \bibnamefont
  {Warren}},\ }\href {\doibase 10.1103/PhysRevLett.87.225702} {\bibfield
  {journal} {\bibinfo  {journal} {Phys. Rev. Lett.}\ }\textbf {\bibinfo
  {volume} {87}},\ \bibinfo {pages} {225702} (\bibinfo {year}
  {2001})}\BibitemShut {NoStop}%
\bibitem [{\citenamefont {Warren}(2003)}]{Warren_PRE_2003}%
  \BibitemOpen
  \bibfield  {author} {\bibinfo {author} {\bibfnamefont {P.~B.}\ \bibnamefont
  {Warren}},\ }\href {\doibase 10.1103/PhysRevE.68.066702} {\bibfield
  {journal} {\bibinfo  {journal} {Phys. Rev. E}\ }\textbf {\bibinfo {volume}
  {68}},\ \bibinfo {pages} {066702} (\bibinfo {year} {2003})}\BibitemShut
  {NoStop}%
\bibitem [{\citenamefont {Vanya}\ \emph {et~al.}(2018)\citenamefont {Vanya},
  \citenamefont {Crout}, \citenamefont {Sharman},\ and\ \citenamefont
  {Elliott}}]{Vanya_PRE_2018}%
  \BibitemOpen
  \bibfield  {author} {\bibinfo {author} {\bibfnamefont {P.}~\bibnamefont
  {Vanya}}, \bibinfo {author} {\bibfnamefont {P.}~\bibnamefont {Crout}},
  \bibinfo {author} {\bibfnamefont {J.}~\bibnamefont {Sharman}}, \ and\
  \bibinfo {author} {\bibfnamefont {J.~A.}\ \bibnamefont {Elliott}},\ }\href
  {\doibase 10.1103/PhysRevE.98.033310} {\bibfield  {journal} {\bibinfo
  {journal} {Phys. Rev. E}\ }\textbf {\bibinfo {volume} {98}},\ \bibinfo
  {pages} {033310} (\bibinfo {year} {2018})}\BibitemShut {NoStop}%
\bibitem [{\citenamefont {Chang}\ \emph {et~al.}(2016)\citenamefont {Chang},
  \citenamefont {Li}, \citenamefont {Li},\ and\ \citenamefont
  {Karniadakis}}]{Chang_PLOS_2016}%
  \BibitemOpen
  \bibfield  {author} {\bibinfo {author} {\bibfnamefont {H.-Y.}\ \bibnamefont
  {Chang}}, \bibinfo {author} {\bibfnamefont {X.}~\bibnamefont {Li}}, \bibinfo
  {author} {\bibfnamefont {H.}~\bibnamefont {Li}}, \ and\ \bibinfo {author}
  {\bibfnamefont {G.~E.}\ \bibnamefont {Karniadakis}},\ }\href {\doibase
  10.1371/journal.pcbi.1005173} {\bibfield  {journal} {\bibinfo  {journal}
  {PLOS Computational Biology}\ }\textbf {\bibinfo {volume} {12}},\ \bibinfo
  {pages} {1} (\bibinfo {year} {2016})}\BibitemShut {NoStop}%
\bibitem [{\citenamefont {Li}\ \emph {et~al.}(2014)\citenamefont {Li},
  \citenamefont {Peng}, \citenamefont {Lei}, \citenamefont {Dao},\ and\
  \citenamefont {Karniadakis}}]{Li_PTRSA_2014}%
  \BibitemOpen
  \bibfield  {author} {\bibinfo {author} {\bibfnamefont {X.}~\bibnamefont
  {Li}}, \bibinfo {author} {\bibfnamefont {Z.}~\bibnamefont {Peng}}, \bibinfo
  {author} {\bibfnamefont {H.}~\bibnamefont {Lei}}, \bibinfo {author}
  {\bibfnamefont {M.}~\bibnamefont {Dao}}, \ and\ \bibinfo {author}
  {\bibfnamefont {G.~E.}\ \bibnamefont {Karniadakis}},\ }\href {\doibase
  10.1098/rsta.2013.0389} {\bibfield  {journal} {\bibinfo  {journal}
  {Philosophical Transactions of the Royal Society A: Mathematical, Physical
  and Engineering Sciences}\ }\textbf {\bibinfo {volume} {372}},\ \bibinfo
  {pages} {20130389} (\bibinfo {year} {2014})}\BibitemShut {NoStop}%
\bibitem [{\citenamefont {Peng}\ \emph {et~al.}(2013)\citenamefont {Peng},
  \citenamefont {Li}, \citenamefont {Pivkin}, \citenamefont {Dao},
  \citenamefont {Karniadakis},\ and\ \citenamefont {Suresh}}]{Peng_PNAS_2013}%
  \BibitemOpen
  \bibfield  {author} {\bibinfo {author} {\bibfnamefont {Z.}~\bibnamefont
  {Peng}}, \bibinfo {author} {\bibfnamefont {X.}~\bibnamefont {Li}}, \bibinfo
  {author} {\bibfnamefont {I.~V.}\ \bibnamefont {Pivkin}}, \bibinfo {author}
  {\bibfnamefont {M.}~\bibnamefont {Dao}}, \bibinfo {author} {\bibfnamefont
  {G.~E.}\ \bibnamefont {Karniadakis}}, \ and\ \bibinfo {author} {\bibfnamefont
  {S.}~\bibnamefont {Suresh}},\ }\href {\doibase 10.1073/pnas.1311827110}
  {\bibfield  {journal} {\bibinfo  {journal} {Proceedings of the National
  Academy of Sciences}\ }\textbf {\bibinfo {volume} {110}},\ \bibinfo {pages}
  {13356} (\bibinfo {year} {2013})},\ \Eprint
  {http://arxiv.org/abs/https://www.pnas.org/content/110/33/13356.full.pdf}
  {https://www.pnas.org/content/110/33/13356.full.pdf} \BibitemShut {NoStop}%
\bibitem [{\citenamefont {Li}\ \emph {et~al.}(2009)\citenamefont {Li},
  \citenamefont {Pivkin}, \citenamefont {Liang},\ and\ \citenamefont
  {Karniadakis}}]{Li_MA_2009}%
  \BibitemOpen
  \bibfield  {author} {\bibinfo {author} {\bibfnamefont {X.}~\bibnamefont
  {Li}}, \bibinfo {author} {\bibfnamefont {I.~V.}\ \bibnamefont {Pivkin}},
  \bibinfo {author} {\bibfnamefont {H.}~\bibnamefont {Liang}}, \ and\ \bibinfo
  {author} {\bibfnamefont {G.~E.}\ \bibnamefont {Karniadakis}},\ }\href
  {\doibase 10.1021/ma9000918} {\bibfield  {journal} {\bibinfo  {journal}
  {Macromolecules}\ }\textbf {\bibinfo {volume} {42}},\ \bibinfo {pages} {3195}
  (\bibinfo {year} {2009})}\BibitemShut {NoStop}%
\bibitem [{\citenamefont {Li}\ \emph {et~al.}(2012)\citenamefont {Li},
  \citenamefont {Li}, \citenamefont {Li},\ and\ \citenamefont
  {Gao}}]{Li_Nanoscale_2012}%
  \BibitemOpen
  \bibfield  {author} {\bibinfo {author} {\bibfnamefont {Y.}~\bibnamefont
  {Li}}, \bibinfo {author} {\bibfnamefont {X.}~\bibnamefont {Li}}, \bibinfo
  {author} {\bibfnamefont {Z.}~\bibnamefont {Li}}, \ and\ \bibinfo {author}
  {\bibfnamefont {H.}~\bibnamefont {Gao}},\ }\href {\doibase
  10.1039/C2NR30379E} {\bibfield  {journal} {\bibinfo  {journal} {Nanoscale}\
  }\textbf {\bibinfo {volume} {4}},\ \bibinfo {pages} {3768} (\bibinfo {year}
  {2012})}\BibitemShut {NoStop}%
\bibitem [{\citenamefont {Ghoufi}\ and\ \citenamefont
  {Malfreyt}(2010)}]{Ghoufi_PRE_2010}%
  \BibitemOpen
  \bibfield  {author} {\bibinfo {author} {\bibfnamefont {A.}~\bibnamefont
  {Ghoufi}}\ and\ \bibinfo {author} {\bibfnamefont {P.}~\bibnamefont
  {Malfreyt}},\ }\href {\doibase 10.1103/PhysRevE.82.016706} {\bibfield
  {journal} {\bibinfo  {journal} {Phys. Rev. E}\ }\textbf {\bibinfo {volume}
  {82}},\ \bibinfo {pages} {016706} (\bibinfo {year} {2010})}\BibitemShut
  {NoStop}%
\bibitem [{\citenamefont {Ghoufi}\ and\ \citenamefont
  {Malfreyt}(2011)}]{Ghoufi_PRE_2011}%
  \BibitemOpen
  \bibfield  {author} {\bibinfo {author} {\bibfnamefont {a.}~\bibnamefont
  {Ghoufi}}\ and\ \bibinfo {author} {\bibfnamefont {P.}~\bibnamefont
  {Malfreyt}},\ }\href {\doibase 10.1103/PhysRevE.83.051601} {\bibfield
  {journal} {\bibinfo  {journal} {Physical Review E - Statistical, Nonlinear,
  and Soft Matter Physics}\ }\textbf {\bibinfo {volume} {83}},\ \bibinfo
  {pages} {1} (\bibinfo {year} {2011})}\BibitemShut {NoStop}%
\bibitem [{\citenamefont {Ghoufi}\ \emph {et~al.}(2013)\citenamefont {Ghoufi},
  \citenamefont {Emile},\ and\ \citenamefont {Malfreyt}}]{Ghoufi_EPJ_2013}%
  \BibitemOpen
  \bibfield  {author} {\bibinfo {author} {\bibfnamefont {A.}~\bibnamefont
  {Ghoufi}}, \bibinfo {author} {\bibfnamefont {J.}~\bibnamefont {Emile}}, \
  and\ \bibinfo {author} {\bibfnamefont {P.}~\bibnamefont {Malfreyt}},\ }\href
  {\doibase 10.1140/epje/i2013-13010-7} {\bibfield  {journal} {\bibinfo
  {journal} {The European physical journal. E, Soft matter}\ }\textbf {\bibinfo
  {volume} {36}},\ \bibinfo {pages} {10} (\bibinfo {year} {2013})}\BibitemShut
  {NoStop}%
\bibitem [{\citenamefont {Merabia}\ and\ \citenamefont
  {Pagonabarraga}(2006)}]{Merabia_EPJE_2006}%
  \BibitemOpen
  \bibfield  {author} {\bibinfo {author} {\bibfnamefont {S.}~\bibnamefont
  {Merabia}}\ and\ \bibinfo {author} {\bibfnamefont {I.}~\bibnamefont
  {Pagonabarraga}},\ }\href {\doibase 10.1140/epje/i2005-10128-1} {\bibfield
  {journal} {\bibinfo  {journal} {The European Physical Journal E}\ }\textbf
  {\bibinfo {volume} {20}},\ \bibinfo {pages} {209} (\bibinfo {year}
  {2006})}\BibitemShut {NoStop}%
\bibitem [{\citenamefont {Merabia}\ \emph {et~al.}(2008)\citenamefont
  {Merabia}, \citenamefont {Bonet-Avalos},\ and\ \citenamefont
  {Pagonabarraga}}]{Merabia_JNNFM_2008}%
  \BibitemOpen
  \bibfield  {author} {\bibinfo {author} {\bibfnamefont {S.}~\bibnamefont
  {Merabia}}, \bibinfo {author} {\bibfnamefont {J.}~\bibnamefont
  {Bonet-Avalos}}, \ and\ \bibinfo {author} {\bibfnamefont {I.}~\bibnamefont
  {Pagonabarraga}},\ }\href {\doibase
  https://doi.org/10.1016/j.jnnfm.2008.01.009} {\bibfield  {journal} {\bibinfo
  {journal} {Journal of Non-Newtonian Fluid Mechanics}\ }\textbf {\bibinfo
  {volume} {154}},\ \bibinfo {pages} {13 } (\bibinfo {year}
  {2008})}\BibitemShut {NoStop}%
\bibitem [{\citenamefont {Merabia}\ and\ \citenamefont
  {Avalos}(2008)}]{Merabia_PRL_2008}%
  \BibitemOpen
  \bibfield  {author} {\bibinfo {author} {\bibfnamefont {S.}~\bibnamefont
  {Merabia}}\ and\ \bibinfo {author} {\bibfnamefont {J.~B.}\ \bibnamefont
  {Avalos}},\ }\href {\doibase 10.1103/PhysRevLett.101.208304} {\bibfield
  {journal} {\bibinfo  {journal} {Phys. Rev. Lett.}\ }\textbf {\bibinfo
  {volume} {101}},\ \bibinfo {pages} {208304} (\bibinfo {year}
  {2008})}\BibitemShut {NoStop}%
\bibitem [{\citenamefont {Espa\~nol}\ and\ \citenamefont
  {Revenga}(2003)}]{Espanol_PRE_2003}%
  \BibitemOpen
  \bibfield  {author} {\bibinfo {author} {\bibfnamefont {P.}~\bibnamefont
  {Espa\~nol}}\ and\ \bibinfo {author} {\bibfnamefont {M.}~\bibnamefont
  {Revenga}},\ }\href {\doibase 10.1103/PhysRevE.67.026705} {\bibfield
  {journal} {\bibinfo  {journal} {Phys. Rev. E}\ }\textbf {\bibinfo {volume}
  {67}},\ \bibinfo {pages} {026705} (\bibinfo {year} {2003})}\BibitemShut
  {NoStop}%
\bibitem [{\citenamefont {Thieulot}\ \emph
  {et~al.}(2005{\natexlab{a}})\citenamefont {Thieulot}, \citenamefont
  {Janssen},\ and\ \citenamefont {Espa\~nol}}]{Thieulot_PRE_2005a}%
  \BibitemOpen
  \bibfield  {author} {\bibinfo {author} {\bibfnamefont {C.}~\bibnamefont
  {Thieulot}}, \bibinfo {author} {\bibfnamefont {L.~P. B.~M.}\ \bibnamefont
  {Janssen}}, \ and\ \bibinfo {author} {\bibfnamefont {P.}~\bibnamefont
  {Espa\~nol}},\ }\href {\doibase 10.1103/PhysRevE.72.016713} {\bibfield
  {journal} {\bibinfo  {journal} {Phys. Rev. E}\ }\textbf {\bibinfo {volume}
  {72}},\ \bibinfo {pages} {016713} (\bibinfo {year}
  {2005}{\natexlab{a}})}\BibitemShut {NoStop}%
\bibitem [{\citenamefont {Thieulot}\ \emph
  {et~al.}(2005{\natexlab{b}})\citenamefont {Thieulot}, \citenamefont
  {Janssen},\ and\ \citenamefont {Espa\~nol}}]{Thieulot_PRE_2005b}%
  \BibitemOpen
  \bibfield  {author} {\bibinfo {author} {\bibfnamefont {C.}~\bibnamefont
  {Thieulot}}, \bibinfo {author} {\bibfnamefont {L.~P. B.~M.}\ \bibnamefont
  {Janssen}}, \ and\ \bibinfo {author} {\bibfnamefont {P.}~\bibnamefont
  {Espa\~nol}},\ }\href {\doibase 10.1103/PhysRevE.72.016714} {\bibfield
  {journal} {\bibinfo  {journal} {Phys. Rev. E}\ }\textbf {\bibinfo {volume}
  {72}},\ \bibinfo {pages} {016714} (\bibinfo {year}
  {2005}{\natexlab{b}})}\BibitemShut {NoStop}%
\bibitem [{\citenamefont {Litvinov}\ \emph {et~al.}(2008)\citenamefont
  {Litvinov}, \citenamefont {Ellero}, \citenamefont {Hu},\ and\ \citenamefont
  {Adams}}]{Litvinov_PRE_2008}%
  \BibitemOpen
  \bibfield  {author} {\bibinfo {author} {\bibfnamefont {S.}~\bibnamefont
  {Litvinov}}, \bibinfo {author} {\bibfnamefont {M.}~\bibnamefont {Ellero}},
  \bibinfo {author} {\bibfnamefont {X.}~\bibnamefont {Hu}}, \ and\ \bibinfo
  {author} {\bibfnamefont {N.~A.}\ \bibnamefont {Adams}},\ }\href {\doibase
  10.1103/PhysRevE.77.066703} {\bibfield  {journal} {\bibinfo  {journal} {Phys.
  Rev. E}\ }\textbf {\bibinfo {volume} {77}},\ \bibinfo {pages} {066703}
  (\bibinfo {year} {2008})}\BibitemShut {NoStop}%
\bibitem [{Note1()}]{Note1}%
  \BibitemOpen
  \bibinfo {note} {In general, any power of the local density can be considered
  and thus the equation of state can be influenced, as was demonstrated by
  Trofimov \protect \emph {et al.}~\cite {Trofimov_JCP_2002}.}\BibitemShut
  {Stop}%
\bibitem [{Note2()}]{Note2}%
  \BibitemOpen
  \bibinfo {note} {There is also a trivial function where the self-energy
  contains only the local density of the particles of the like type,
  $u[\protect \mathaccentV {bar}016\rho ^{(\alpha )}(\protect \mathbf {
  r}_i)]$. However, the resulting force is the same as for the partial local
  density variant.}\BibitemShut {Stop}%
\bibitem [{\citenamefont {Seaton}\ \emph {et~al.}(2013)\citenamefont {Seaton},
  \citenamefont {Anderson}, \citenamefont {Metz},\ and\ \citenamefont
  {Smith}}]{DLMS}%
  \BibitemOpen
  \bibfield  {author} {\bibinfo {author} {\bibfnamefont {M.~A.}\ \bibnamefont
  {Seaton}}, \bibinfo {author} {\bibfnamefont {R.~L.}\ \bibnamefont
  {Anderson}}, \bibinfo {author} {\bibfnamefont {S.}~\bibnamefont {Metz}}, \
  and\ \bibinfo {author} {\bibfnamefont {W.}~\bibnamefont {Smith}},\ }\href
  {\doibase 10.1080/08927022.2013.772297} {\bibfield  {journal} {\bibinfo
  {journal} {Molecular Simulation}\ }\textbf {\bibinfo {volume} {39}},\
  \bibinfo {pages} {796} (\bibinfo {year} {2013})},\ \Eprint
  {http://arxiv.org/abs/http://dx.doi.org/10.1080/08927022.2013.772297}
  {http://dx.doi.org/10.1080/08927022.2013.772297} \BibitemShut {NoStop}%
\bibitem [{\citenamefont {Warren}(2013)}]{Warren_PRE_2013}%
  \BibitemOpen
  \bibfield  {author} {\bibinfo {author} {\bibfnamefont {P.~B.}\ \bibnamefont
  {Warren}},\ }\href {\doibase 10.1103/PhysRevE.87.045303} {\bibfield
  {journal} {\bibinfo  {journal} {Phys. Rev. E}\ }\textbf {\bibinfo {volume}
  {87}},\ \bibinfo {pages} {045303} (\bibinfo {year} {2013})}\BibitemShut
  {NoStop}%
\bibitem [{\citenamefont {Vanya}\ \emph {et~al.}(2019)\citenamefont {Vanya},
  \citenamefont {Sharman},\ and\ \citenamefont {Elliott}}]{Vanya_JCP_2019}%
  \BibitemOpen
  \bibfield  {author} {\bibinfo {author} {\bibfnamefont {P.}~\bibnamefont
  {Vanya}}, \bibinfo {author} {\bibfnamefont {J.}~\bibnamefont {Sharman}}, \
  and\ \bibinfo {author} {\bibfnamefont {J.~A.}\ \bibnamefont {Elliott}},\
  }\href {\doibase 10.1063/1.5046851} {\bibfield  {journal} {\bibinfo
  {journal} {The Journal of Chemical Physics}\ }\textbf {\bibinfo {volume}
  {150}},\ \bibinfo {pages} {064101} (\bibinfo {year} {2019})}\BibitemShut
  {NoStop}%
\bibitem [{\citenamefont {Rafii-Tabar}\ and\ \citenamefont
  {Sulton}(1991)}]{RafiiTabar_PML_1991}%
  \BibitemOpen
  \bibfield  {author} {\bibinfo {author} {\bibfnamefont {H.}~\bibnamefont
  {Rafii-Tabar}}\ and\ \bibinfo {author} {\bibfnamefont {A.~P.}\ \bibnamefont
  {Sulton}},\ }\href {\doibase 10.1080/09500839108205994} {\bibfield  {journal}
  {\bibinfo  {journal} {Philosophical Magazine Letters}\ }\textbf {\bibinfo
  {volume} {63}},\ \bibinfo {pages} {217} (\bibinfo {year} {1991})}\BibitemShut
  {NoStop}%
\bibitem [{\citenamefont {Johnson}(1989)}]{Johnson_PRB_1989}%
  \BibitemOpen
  \bibfield  {author} {\bibinfo {author} {\bibfnamefont {R.~A.}\ \bibnamefont
  {Johnson}},\ }\href {\doibase 10.1103/PhysRevB.39.12554} {\bibfield
  {journal} {\bibinfo  {journal} {Phys. Rev. B}\ }\textbf {\bibinfo {volume}
  {39}},\ \bibinfo {pages} {12554} (\bibinfo {year} {1989})}\BibitemShut
  {NoStop}%
\bibitem [{Note3()}]{Note3}%
  \BibitemOpen
  \bibinfo {note} {Available at \protect \url
  {https://github.com/petervanya/DPDsim}.}\BibitemShut {Stop}%
\end{thebibliography}%



@article{Pagonabarraga_JCP_2001,
author = {Pagonabarraga, I. and Frenkel, D.},
doi = {10.1063/1.1396848},
issn = {00219606},
journal = {The Journal of Chemical Physics},
number = {11},
pages = {5015--5026},
title = {{Dissipative particle dynamics for interacting systems}},
volume = {115},
year = {2001}
}


@article{Ghoufi_EPJ_2013,
abstract = {Many Body Dissipative Particles Dynamics (MDPD) simulation is a novel promising mesoscopic method to model the liquid-vapor interfaces. Based upon works of Paganobarraga and Frenkel (J. Chem. Phys. 15, 5015 (2001)) and Trofimov (J. Chem. Phys. 117, 9383 (2002)) and of Warren (Phys. Rev. E 68, 066702 (2003)) this method has been critically reviewed during this last decade. We propose here to give an overview of the Many Body Dissipative Particles Dynamic simulation within the framework of the liquid-vapor interfaces. We recall the theoretical background of MDPD and we present some recent results of systems of interest such as water liquid-vapor interfaces and salt effect on water surface tension. Additionally we discuss the ability of MDPD to capture the mechanisms at the mesoscopic scale through the formation of micelles and the coalescence of a nanodroplet water on water surface.},
author = {Ghoufi, Aziz and Emile, Janine and Malfreyt, Patrice},
doi = {10.1140/epje/i2013-13010-7},
isbn = {1292-8941},
issn = {1292895X},
journal = {The European physical journal. E, Soft matter},
number = {1},
pages = {10},
pmid = {23361618},
title = {{Recent advances in Many Body Dissipative Particles Dynamics simulations of liquid-vapor interfaces.}},
volume = {36},
year = {2013}
}


@article{Ghoufi_JCTC_2012,
abstract = {Modeling interfacial properties is a major challenge for mesoscopic simulation methods. Many-body dissipative particle dynamics (MDPD) is then a promising method to model heterogeneous systems at long time and length scales. However no rule exists to obtain a set of MDPD parameters capable to reproduce the thermodynamic properties of a molecular system of a specific chemistry. In this letter, we provide a general multiscale method to obtain a set of parameters from atomistic simulations using Flory?Huggins theory (FH) to be used with dissipative particle dynamics. We demonstrate the high quality and the transferability of the resulting parameters on the salt concentration dependence of surface tension. We also show the specificity of inorganic salt at the water?air interface. Our results indicate that the increase of surface tension with the salt concentration cannot be explained in terms of the charge image concept based on the Wagner, Onsager, and Samaras theory but rather in terms of the ion hydration.$\backslash$nModeling interfacial properties is a major challenge for mesoscopic simulation methods. Many-body dissipative particle dynamics (MDPD) is then a promising method to model heterogeneous systems at long time and length scales. However no rule exists to obtain a set of MDPD parameters capable to reproduce the thermodynamic properties of a molecular system of a specific chemistry. In this letter, we provide a general multiscale method to obtain a set of parameters from atomistic simulations using Flory?Huggins theory (FH) to be used with dissipative particle dynamics. We demonstrate the high quality and the transferability of the resulting parameters on the salt concentration dependence of surface tension. We also show the specificity of inorganic salt at the water?air interface. Our results indicate that the increase of surface tension with the salt concentration cannot be explained in terms of the charge image concept based on the Wagner, Onsager, and Samaras theory but rather in terms of the ion hydration.},
author = {Ghoufi, Aziz and Malfreyt, Patrice},
doi = {10.1021/ct200833s},
isbn = {1549-9618},
issn = {15499618},
journal = {Journal of Chemical Theory and Computation},
number = {3},
pages = {787--791},
pmid = {26593339},
title = {{Coarse grained simulations of the electrolytes at the water-air interface from many body dissipative particle dynamics}},
volume = {8},
year = {2012}
}


@article{Ghoufi_PRE_2011,
abstract = {We report a mesoscale modeling of the liquid-vapor interface of water. A mesoscopic model of water has been established in dissipative particle dynamics (DPD) to reproduce the interfacial properties of water. The surface tension and coexisting densities are compared between atomistic and mesoscopic simulations. Simple scaling relations have been established to link the atomistic and mesoscopic length and time scales. Our study demonstrates the capability of the DPD method to explore the interfacial properties of a planar water liquid-vapor interface and a water nanodroplet. This constitutes an important step toward the calculation of the surface tension of larger and more complex interfacial systems.},
author = {Ghoufi, a. and Malfreyt, P.},
doi = {10.1103/PhysRevE.83.051601},
issn = {15393755},
journal = {Physical Review E - Statistical, Nonlinear, and Soft Matter Physics},
number = {5},
pages = {1--5},
pmid = {21728541},
title = {{Mesoscale modeling of the water liquid-vapor interface: A surface tension calculation}},
volume = {83},
year = {2011}
}


@article{Ghoufi_PRE_2010,
  title = {Calculation of the surface tension from multibody dissipative particle dynamics and Monte Carlo methods},
  author = {Ghoufi, A. and Malfreyt, P.},
  journal = {Phys. Rev. E},
  volume = {82},
  issue = {1},
  pages = {016706},
  numpages = {11},
  year = {2010},
  month = {Jul},
  publisher = {American Physical Society},
  doi = {10.1103/PhysRevE.82.016706},
  url = {https://link.aps.org/doi/10.1103/PhysRevE.82.016706}
}


@article{Groot_BiophysJ_2001,
abstract = {A new simulation method, dissipative particle dynamics, is applied to model biological membranes. In this method, several atoms are united into a single simulation particle. The solubility and compressibility of the various liquid components are reproduced by the simulation model. When applied to a bilayer of phosphatidylethanolamine, the membrane structure obtained matches quantitatively with full atomistic simulations and with experiments reported in the literature. The method is applied to investigate the cause of cell death when bacteria are exposed to nonionic surfactants. Mixed bilayers of lipid and nonionic surfactant were studied, and the diffusion of water through the bilayer was monitored. Small transient holes are seen to appear at 40\% mole-fraction C(9)E(8), which become permanent holes between 60 and 70\% surfactant. When C(12)E(6) is applied, permanent holes only arise at 90\% mole-fraction surfactant. Some simulations have been carried out to determine the rupture properties of mixed bilayers of phosphatidylethanolamine and C(12)E(6). These simulations indicate that the area of a pure lipid bilayer can be increased by a factor 2. The inclusion of surfactant considerably reduces both the extensibility and the maximum stress that the bilayer can withstand. This may explain why dividing cells are more at risk than static cells.},
author = {Groot, R D and Rabone, K L},
doi = {10.1016/S0006-3495(01)75737-2},
isbn = {0006-3495},
issn = {00063495},
journal = {Biophysical journal},
number = {2},
pages = {725--736},
pmid = {11463621},
publisher = {Elsevier},
title = {{Mesoscopic simulation of cell membrane damage, morphology change and rupture by nonionic surfactants.}},
url = {http://dx.doi.org/10.1016/S0006-3495(01)75737-2},
volume = {81},
year = {2001}
}


@article{Groot_JCP_1998,
abstract = {The dissipative particle dynamics ͑DPD͒ simulation method has been used to study mesophase formation of linear (A m B n) diblock copolymer melts. The polymers are represented by relatively short strings of soft spheres, connected by harmonic springs. These melts spontaneously form a mesocopically ordered structure, depending on the length ratio of the two blocks and on the Flory–Huggins ␹-parameter. The main emphasis here is on validation of the method and model by comparing the predicted equilibrium phases to existing mean-field theory and to experimental results. The real strength of the DPD method, however, lies in its capability to predict the dynamical pathway along which a block copolymer melt finds its equilibrium structure after a temperature quench. The present work has led to the following results: ͑1͒ As the polymer becomes more asymmetric, we qualitatively find the order of the equilibrium structures as lamellar, perforated lamellar, hexagonal rods, micelles. Qualitatively this is in agreement with experiments and existing mean-field theory. After taking fluctuation corrections to the mean field theory into account, a quantitative match for the locations of the phase transitions is found. ͑2͒ Where mean-field theory predicts the gyroid phase to be stable, the simulations evolve toward the hexagonally perforated lamellar phase. ͑3͒ When a melt is quenched the stable structure emerges via a nontrivial pathway, where a series of metastable phases can be formed before equilibrium is reached. The pathway to equilibrium involves a percolation of the minority phase into a network of tubes, which is destabilized by a nematic or smectic transition. ͑4͒ We conclude that either hydrodynamic interactions, or the precise form of the Onsager kinetic coefficient play an important role in the evolution of the mesophases.},
author = {Groot, Robert D and Madden, Timothy J},
doi = {10.1063/1.1642602},
issn = {00219606},
journal = {The Journal of Chemical Physics},
number = {3949},
pages = {4804608--114904},
pmid = {75255700047},
title = {{Dynamic simulation of diblock copolymer microphase separation}},
url = {http://dx.doi.org/10.1063/1.476300$\backslash$nhttp://dx.doi.org/10.1063/1.476300},
volume = {108},
year = {1998}
}


@article{Groot_JCP_1999,
author = {Robert D. Groot and Timothy J. Madden and Dominic J. Tildesley},
title = {On the role of hydrodynamic interactions in block copolymer microphase separation},
journal = {The Journal of Chemical Physics},
volume = {110},
number = {19},
pages = {9739-9749},
year = {1999},
doi = {10.1063/1.478939},
URL = {http://dx.doi.org/10.1063/1.478939},
eprint = {http://dx.doi.org/10.1063/1.478939}
}


@article{Groot_JCP_2003,
abstract = {Electrostatic interactions have been incorporated in dissipative particle dynamics (DPD) simulation. The electrostatic field is solved locally on a grid. Within this formalism, local inhomogeneities in the electrostatic permittivity can be treated without any problem. Key issues like the screening of the potential near a charged surface and the Stillinger-Lovett moment conditions are satisfied. This implies that the method captures the essential features of electrostatic interaction. For the direct simulation of mixed surfactants near oil-water interfaces, or for the simulation of Coulombic polymer-surfactant interactions, this method has all the advantages of DPD over full atomistic molecular dynamics (MD). DPD has proven to be faster than MD by many orders of magnitude, depending on the precise scaling factor chosen for the simulation. This brings phenomena of microseconds in reach of routine simulation, while maintaining a fairly accurate representation of the structure of the molecules. As an example of this simulation tool, the interaction between a cationic polyelectrolyte and anionic surfactant is discussed. Without a surfactant, the polyelectrolyte shows a fractal dimensionality that is in line with the theoretical and experimental values cited in literature, it behaves as a fairly stiff rod, d(f)similar to1.1. When salt is replaced by anionic surfactant, the polymer wraps around one or more discrete surfactant micelles, in line with the current understanding of these systems, and scaling invariance in the correlation function is broken. (C) 2003 American Institute of Physics.},
author = {Groot, R. D.},
doi = {10.1063/1.1574800},
isbn = {0021-9606},
issn = {00219606},
journal = {Journal of Chemical Physics},
number = {24},
pages = {11265--11277},
pmid = {183402200049},
title = {{Electrostatic interactions in dissipative particle dynamics-simulation of polyelectrolytes and anionic surfactants}},
volume = {118},
year = {2003}
}


@article{Groot_MA_1995,
abstract = {An off-lattice simulation model for associative polymer gels as introduced recently (Groot, R. D.; Agterof, W. G. M. J. Chem. Phys. 1994, 100, 1649) has been applied to obtain the mechanical spectrum as a function of frequency and polymer concentration, using a Green-Kubo relation for the time- dependent modulus. Two stages of relaxation are observable in our simulations. The early-time decay is consistent with a -2/3 power law, whose form is insensitive to large variations in polymer concentration, association lifetime, and degree of association. The late stage, which relaxes like the end-to-end vector in the Rouse model, has a characteristic stress that scales as the cube of the concentration and a relaxation time that is proportional to the monomer-monomer dissociation rate. The simulation results have been compared with experiments found in the literature for several physical gels. The quantitative agreement calls into question other postulated mechanisms involving hydrodynamic interaction or reptation, since the simulation contains neither of these features. As an alternative explanation for the observed early-time decay, an explicit relation between the power law exponent and the polymer fractal dimension is given.},
author = {Groot, Robert D. and Agterof, Wim G. M.},
doi = {10.1021/ma00122a041},
isbn = {0024929719},
issn = {0024-9297},
journal = {Macromolecules},
pages = {6284--6295},
title = {{Dynamic Viscoelastic Modulus of Associative Polymer Networks: Off-Lattice Simulations, Theory and Comparison to Experiments}},
url = {http://pubs.acs.org/doi/abs/10.1021/ma00122a041},
volume = {28},
year = {1995}
}


@article{Hoffmann_JCP_1997,
author = {Hoffmann, Alexander and Sommer, Jens-Uwe and Blumen, Alexander},
doi = {10.1063/1.473668},
issn = {00219606},
journal = {The Journal of Chemical Physics},
number = {16},
pages = {6709},
title = {{Statics and dynamics of dense copolymer melts: A Monte Carlo simulation study}},
url = {http://link.aip.org/link/JCPSA6/v106/i16/p6709/s1\&Agg=doi},
volume = {106},
year = {1997}
}


@article{Hossain_JCP_2010,
abstract = {Molecular dynamics simulations were used to study deformation mechanisms during uniaxial tensile deformation of an amorphous polyethylene polymer. The stress-strain behavior comprised elastic, yield, strain softening and strain hardening regions that were qualitatively in agreement with previous simulations and experimental results. The chain lengths, number of chains, strain rate and temperature dependence of the stress-strain behavior was investigated. The energy contributions from the united atom potential were calculated as a function of strain to help elucidate the inherent deformation mechanisms within the elastic, yield, and strain hardening regions. The results of examining the partitioning of energy show that the elastic and yield regions were mainly dominated by interchain non-bonded interactions whereas strain hardening regions were mainly dominated by intra-chain dihedral motion of polyethylene. Additional results show how internal mechanisms associated with bond length, bond angle, dihedral distributions, change of free volume and chain entanglements evolve with increasing deformation. ?? 2010 Elsevier Ltd.},
author = {Hossain, D. and Tschopp, M. a. and Ward, D. K. and Bouvard, J. L. and Wang, P. and Horstemeyer, M. F.},
doi = {10.1016/j.polymer.2010.10.009},
isbn = {00323861},
issn = {00323861},
journal = {Polymer},
keywords = {Deformation mechanisms,Molecular dynamics simulation,Polyethylene},
number = {25},
pages = {6071--6083},
publisher = {Elsevier Ltd},
title = {{Molecular dynamics simulations of deformation mechanisms of amorphous polyethylene}},
url = {http://dx.doi.org/10.1016/j.polymer.2010.10.009},
volume = {51},
year = {2010}
}


@article{Bartlett_PRL_1992,
abstract = {We report a detailed experimental study of the superlattice structures formed in dense binary mixtures of hard-sphere colloids. The phase diagrams observed depend sensitively on the ratio alpha=R(S)/R(L) of the radii of the small (S) and large (L) components. Mixtures of size ratio alpha=0.72, 0.52, 0.42, and 0.39 are studied. The structures of the colloidal phases formed were identified using a combination of light-scattering techniques and confocal fluorescent microscopy. At alpha=0.39, ordered binary crystals are formed in suspensions containing an equal number of large and small spheres which microscopy shows have a three-dimensional structure similar to either NaCl or NiAs. At the larger size ratio, alpha=0.52, we observe LS2 and LS13 superlattices, isostructural to the molecular compounds AlB2 and NaZn13, while at alpha=0.72 the two components are immiscible in the solid state and no superlattice structures are found. These experimental observations are compared with the predictions of Monte Carlo simulations and cell model theories.},
author = {Bartlett, P. and Ottewill, R. H. and Pusey, P. N.},
doi = {10.1103/PhysRevLett.68.3801},
isbn = {1063-651X},
issn = {00319007},
journal = {Physical Review Letters},
number = {25},
pages = {3801--3804},
pmid = {11088547},
title = {{Superlattice formation in binary mixtures of hard-sphere colloids}},
volume = {68},
year = {1992}
}


@article{Hwang_JES_2009,
abstract = {We show a criticality of three water morphological transitions on pore-water transport and proton conductivity in Nafion of a polymer electrolyte membrane fuel cell, addressing its pore-size distribution and the Schr\"{o}der paradox. The first transition leads to the onset of proton conductivity; the second allows for the onset of the capillary percolation channels and proton conductivity jump at a low water content $\lambda$ im , H 2 O = 5 . Using this as immobile saturation, the predicted water distribution and cell performance are in reasonable agreement with the available experiments. The third (the paradox, postulating further capillary advancing) bounds the maximum water content on the cathode side of Nafion, which is also supported by the proposed adsorption isobar (thermodynamic equilibrium limit). These transitions appear in the available pore-size experiments which show capillary percolation channel sizes. In addition, the optimal Nafion pore-water content is between the second and third transitions.},
author = {Hwang, G S and Kaviany, M and Nam, J H and Kim, M H and Son, S Y},
doi = {10.1149/1.3186027},
isbn = {0013-4651},
issn = {00134651},
journal = {Journal of the Electrochemical Society},
number = {10},
pages = {B1192--B1200},
title = {{Pore-Water Morphological Transitions in Polymer Electrolyte of a Fuel Cell}},
url = {http://jes.ecsdl.org/content/156/10/B1192.abstract},
volume = {156},
year = {2009}
}


@article{Izvekov_JCP_2005,
abstract = {The solvation and transport of the hydrated excess proton is studied using the Car-Parrinello molecular-dynamics (CPMD) simulation method. The simulations were performed using BLYP and HCTH gradient-corrected exchange-correlation energy functionals. The fictitious electronic mass was chosen to be small enough so that the underlying water structural and dynamical properties were converged with respect to this important CPMD simulation parameter. An unphysical overstructuring of liquid water in the CPMD simulations using the BLYP functional resulted in the formation of long-lived hydrogen-bonding structures involving the excess proton and a particular (special) water oxygen. The excess proton was observed to be attracted to the special oxygen through the entire length of the BLYP CPMD simulations. Consequently, the excess proton diffusion was limited by the mobility of the special oxygen in the slowly diffusing water network and, in turn, the excess proton self-diffusion coefficient was found to be significantly below the experimental value. On the other hand, the structural properties of liquid water in the HCTH CPMD simulation were seen to be in better agreement with experiment, although the water and excess proton diffusions were still well below the experimental value.},
author = {Izvekov, Sergei and Voth, Gregory a.},
doi = {10.1063/1.1961443},
isbn = {0021-9606},
issn = {00219606},
journal = {Journal of Chemical Physics},
number = {4},
pmid = {16095367},
title = {{Ab initio molecular-dynamics simulation of aqueous proton solvation and transport revisited}},
volume = {123},
year = {2005}
}


@article{Johnson_MolPhys_1993,
abstract = {We review the existing simulation data and equations of state for the Lennard- Jones (L J) fluid, and present new simulation results for both the cut and shifted and the full LJ potential. New parameters for the modified Benedict- Webb-Rubin (MBWR) equation of state used by Nicolas, Gubbins, Streett and Tildesley are presented. In contrast to previous equations, the new equation is accurate for calculations of vapour-liquid equilibria. The equation also accurately correlates pressures and internal energies from the triple point to about 4.5 times the critical temperature over the entire fluid range. An equation of state for the cut and shifted LJ fluid is presented and compared with the simulation data of this work, and previously published Gibbs ensemble data. The MBWR equation of state can be extended to mixtures via the van der Waals one-fluid theory mixing rules. Calculations for binary fluid mixtures are found to be accurate when compared with Gibbs ensemble simulations.},
author = {Johnson, J. Karl and Zollweg, John a. and Gubbins, Keith E.},
doi = {10.1080/00268979300100411},
isbn = {0026-8976},
issn = {0026-8976},
journal = {Molecular Physics},
number = {3},
pages = {591--618},
title = {{The Lennard-Jones equation of state revisited}},
volume = {78},
year = {1993}
}


@article{Kacar_EPL_2013,
abstract = {We present a generalized protocol to compute dissipative particle dynamics interaction parameters where beads may have variable (local) densities in the simulations. A generalized relationship for pair-wise interactions is derived and proof-of-concept simulations are performed. In this general relation, excess repulsions are related to the experimental Flory-Huggins parameters for any molar volume ratio of beads.},
author = {Kacar, G. and Peters, E. a. J. F. and de With, G.},
doi = {10.1209/0295-5075/102/40009},
issn = {0295-5075},
journal = {EPL (Europhysics Letters)},
number = {4},
pages = {40009},
title = {{A generalized method for parameterization of dissipative particle dynamics for variable bead volumes}},
url = {http://stacks.iop.org/0295-5075/102/i=4/a=40009?key=crossref.5c6cfaca878aa95570b8faac409660bb},
volume = {102},
year = {2013}
}


@article{Kinjo_MolSim_2007,
abstract = {To reduce computational cost in large scale molecular$\backslash$nsimulations and to adjust the simulation methods to multiscale nature$\backslash$nof complex materials, it is effective to treat several atoms (or$\backslash$nmolecules) as one element. Dissipative particle dynamics (DPD) and$\backslash$nBrownian dynamics (BD) simulations are typical examples of such$\backslash$ncoarse-graining methods. In the coarse-grained (CG) simulation$\backslash$nmethods, linkage between molecular and mesoscale parameters is$\backslash$nimportant to assess accuracy and applicability of these methods. For$\backslash$nthat purpose, we derived equation of motion for the CG particles by$\backslash$nusing projection operator method which will be appeared on a$\backslash$nsubsequent paper. In the derived equation, the force acting on the CG$\backslash$nparticles is divided into the mean force, friction force and random$\backslash$nforce. In this study, we calculated the mean force between CG$\backslash$nparticles by molecular dynamics (MD) simulations with constraints. We$\backslash$nalso showed the universality of the calculated mean forces. (15$\backslash$nRefs.)},
author = {Kinjo, T and Hyodo, S},
doi = {10.1080/08927020601155436},
isbn = {0892702060},
issn = {0892-7022},
journal = {Molecular Simulation},
keywords = {coarse-graining,mean force,molecular dynamics simulation,projection operator},
number = {4},
pages = {417--420},
title = {{Linkage between atomistic and mesoscale coarse-grained simulation}},
url = {papers://078f6c0c-8070-48cf-a26f-997df9318c5a/Paper/p53},
volume = {33},
year = {2007}
}


@article{Kinjo_PRE_2007,
abstract = {We have derived an equation of motion for coarse-grained particles by using a projection operator. Because the derived coarse-grained equation is based on microscopic description, it can be the basis for models of various coarse-grained simulations. We show that by substitution of random forces into fluctuating forces in the coarse-grained equation, the equations for Brownian dynamics and dissipative particle dynamics are reproduced.},
author = {Kinjo, Tomoyuki and Hyodo, Shi Aki},
doi = {10.1103/PhysRevE.75.051109},
issn = {15393755},
journal = {Physical Review E - Statistical, Nonlinear, and Soft Matter Physics},
number = {5},
pages = {1--9},
pmid = {17677024},
title = {{Equation of motion for coarse-grained simulation based on microscopic description}},
volume = {75},
year = {2007}
}


@article{Lee_JCP_2016,
author = {Lee, Ming Tsung and Vishnyakov, Aleksey and Neimark, Alexander V.},
doi = {10.1063/1.4938271},
isbn = {1848445083},
issn = {00219606},
journal = {Journal of Chemical Physics},
number = {1},
title = {{Coarse-grained model of water diffusion and proton conductivity in hydrated polyelectrolyte membrane}},
url = {http://dx.doi.org/10.1063/1.4938271},
volume = {144},
year = {2016}
}


@article{Lee_JCP_2015,
abstract = {We suggest a coarse-grained model for dissipative particle dynamics (DPD) simulations of solutions with dissociated protons. The model uses standard short-range soft repulsion and smeared charge electrostatic potentials between the beads, representing solution components. The proton is introduced as a separate charged bead that forms dissociable bonds with proton receptor base beads, such as water or deprotonated acid anions. The proton–base bonds are described by Morse potentials. When the proton establishes the Morse bonds with two bases, they form an intermediate complex, and the proton is able to “hop” between the bases artificially mimicking the Grotthuss diffusion mechanism. By adjusting the Morse potential parameters, one can regulate the potential barrier associated with intermediate complex formation and breakup and control the hopping frequency. This makes the proposed model applicable to simulations of proton mobility and reaction equilibria between protonated and deprotonated acid forms in aqu...},
author = {Lee, Ming Tsung and Vishnyakov, Aleksey and Neimark, Alexander V.},
doi = {10.1021/acs.jctc.5b00467},
issn = {15499626},
journal = {Journal of Chemical Theory and Computation},
number = {9},
pages = {4395--4403},
title = {{Modeling Proton Dissociation and Transfer Using Dissipative Particle Dynamics Simulation}},
volume = {11},
year = {2015}
}


@article{Lei_PRE_2010,
abstract = {Starting from microscopic molecular-dynamics (MD) simulations of constrained Lennard-Jones (LJ) clusters (with constant radius of gyration R(g)), we construct two mesoscopic models [Langevin dynamics and dissipative particle dynamics (DPD)] by coarse graining the LJ clusters into single particles. Both static and dynamic properties of the coarse-grained models are investigated and compared with the MD results. The effective mean force field is computed as a function of the intercluster distance, and the corresponding potential scales linearly with the number of particles per cluster and the temperature. We verify that the mean force field can reproduce the equation of state of the atomistic systems within a wide density range but the radial distribution function only within the dilute and the semidilute regime. The friction force coefficients for both models are computed directly from the time-correlation function of the random force field of the microscopic system. For high density or a large cluster size the friction force is overestimated and the diffusivity underestimated due to the omission of many-body effects as a result of the assumed pairwise form of the coarse-grained force field. When the many-body effect is not as pronounced (e.g., smaller R(g) or semidilute system), the DPD model can reproduce the dynamic properties of the MD system.},
author = {Lei, Huan and Caswell, Bruce and Karniadakis, George Em},
doi = {10.1103/PhysRevE.81.026704},
issn = {15393755},
journal = {Physical Review E - Statistical, Nonlinear, and Soft Matter Physics},
number = {2},
pages = {1--10},
pmid = {20365672},
title = {{Direct construction of mesoscopic models from microscopic simulations}},
volume = {81},
year = {2010}
}


@article{Leibler_MA_1980,
abstract = {A microscopic statistical theory of phase equilibria in noncrystalline block copolymers of type AB is developed. In particular, the onset of an ordered mesophase from a homogeneous melt is studied and a criterion of the microphase separation is found. The ...$\backslash$n},
archivePrefix = {arXiv},
arxivId = {arXiv:cond-mat/0402594v3},
author = {Leibler, L},
doi = {10.1021/ma60078a047},
eprint = {0402594v3},
isbn = {0024-9297},
issn = {0024-9297},
journal = {Macromolecules},
number = {10},
pages = {1602--1617},
primaryClass = {arXiv:cond-mat},
title = {{Theory of microphase separation in block copolymers}},
url = {http://pubs.acs.org/doi/abs/10.1021/ma60078a047},
volume = {13},
year = {1980}
}



@article{Li_Plos_2016,
author = {Li, Xiaoxu and Gao, Lianghui and Fang, Weihai},
doi = {10.1371/journal.pone.0154568},
issn = {19326203},
journal = {PLoS ONE},
number = {5},
pages = {26--28},
title = {{Dissipative particle dynamics simulations for phospholipid membranes based on a four-to-one coarse-grained mapping scheme}},
volume = {11},
year = {2016}
}


@article{Lyubartsev_CPL_2000,
abstract = {We present a novel approach to construct pairwise site–site intermolecular interaction potentials from the results of ab initio quantum-mechanical modeling to be used in classical molecular simulations. In the present application, as a first step, we have carried out ab initio Car–Parrinello molecular dynamics (MD) simulations of liquid water to calculate the radial distribution functions (RDF) between all the pairs of nuclei. A second step involves use of the inverse Monte Carlo method, recently developed by us [Phys. Rev. E 52 (199). 3730], to construct a full set of site–site interaction potentials from the ab initio RDFs. The resulting three-site SPC-like water model, which reproduces the liquid structure obtained in the Car–Parrinello MD simulations, can be used as a potential function in classical molecular simulations. We have also found that there exist no pair potential for a rigid three-site water model which exactly satisfy the experimental water RDFs.},
author = {Lyubartsev, a P and Laaksonen, a},
doi = {10.1016/S0009-2614(00)00592-3},
issn = {00092614},
journal = {Chemical Physics Letters},
number = {July},
pages = {15--21},
title = {{Determination of effective pair potentials from ab initio simulations : application to liquid water}},
year = {2000}
}


@article{Lyubartsev_PRE_1995,
abstract = {An approach is presented to solve the reverse problem of statistical mechanics: reconstruction of interaction potentials from radial distribution functions. The method consists of the iterative adjustment of the interaction potential to known radial distribution functions using a Monte Carlo simulation technique and statistical-mechanics relations to connect deviations of canonical averages with Hamiltonian parameters. The method is applied to calculate the effective interaction potentials between the ions in aqueous NaCl solutions at two different concentrations. The reference ion-ion radial distribution functions, calculated in separate molecular dynamics simulations with water molecules, are reproduced in Monte Carlo simulations, using the effective interaction potentials for the hydrated ions. Application of the present method should provide an effective and economical way to simulate equilibrium properties for very large molecular systems (e.g., polyelectrolytes) in the presence of hydrated ions, as well as to offer an approach to reduce a complexity in studies of various associated and aggregated systems in solution.},
author = {Lyubartsev, Alexander P. and Laaksonen, Aatto},
doi = {10.1103/PhysRevE.52.3730},
issn = {1063651X},
journal = {Physical Review E},
number = {4},
pages = {3730--3737},
pmid = {9963851},
title = {{Calculation of effective interaction potentials from radial distribution functions: A reverse Monte Carlo approach}},
volume = {52},
year = {1995}
}


@article{Matsen_PRL_1994,
abstract = {Using self-consistent Geld theory, we examine mierophases of diblock$\backslash$ncopolymers and Gnd, in$\backslash$n$\backslash$naddition to lamellar, hexagonal, and cubic phases, a stable gyroid$\backslash$nphase which occurs between the$\backslash$n$\backslash$nlamellar and hexagonal ones. It terminates at a triple point, with$\backslash$na lamellar to hexagonal transition$\backslash$n$\backslash$noccurring in the weak-segregation limit. Other phases of experimental$\backslash$ninterest are studied, and we$\backslash$n$\backslash$ndescribe the regions in which they are most nearly stable.},
author = {Matsen, M. W. and Schick, M.},
doi = {10.1103/PhysRevLett.72.2660},
isbn = {0031-9007},
issn = {00319007},
journal = {Physical Review Letters},
number = {16},
pages = {2660--2663},
pmid = {10055940},
title = {{Stable and unstable phases of a diblock copolymer melt}},
volume = {72},
year = {1994}
}


@article{Peter_JCP_2014,
abstract = {We present a polarizable water model for the Dissipative Particle Dynamics (DPD) method. Employing long-range electrostatics and Drude oscillators, we calibrate the model using the compressibility and the dielectric constant of water. We validate the model by sampling the dielectric properties of solutions of sodium chloride at various concentrations. Additionally, we apply our model in equilibrium and electroporation simulations of a pure dipalmitoylphosphatidylcholine (DPPC) bilayer, a pure cholesterol domain and a mixed DPPC-cholesterol membrane in polarizable water. Finally, we simulate the transport of a short DNA segment through a DPPC bilayer driven by an external electric field. The new water model is suitable for the DPD simulations of systems where polarization effects play an essential role.},
author = {Peter, Emanuel K. and Pivkin, Igor V.},
doi = {10.1063/1.4899317},
isbn = {doi:10.1063/1.4899317},
issn = {00219606},
journal = {Journal of Chemical Physics},
number = {16},
pmid = {25362324},
title = {{A polarizable coarse-grained water model for dissipative particle dynamics}},
url = {http://dx.doi.org/10.1063/1.4899317},
volume = {141},
year = {2014}
}


@article{Peter_Faraday_2010,
abstract = {This paper gives a short introduction to multiscale simulation approaches in soft matter science. This paper is based on and extended from a previous review. (1. C. Peter and K. Kremer, Soft Matter, 2009, DOI:10.1039/b912027k.) It also includes a discussion of aspects of soft matter in general and a short account of one of the historically underlying concepts, namely renormalization group theory. Some different concepts and several typical problems are shortly addressed, including a (more personal) view on challenges and chances.},
author = {Peter, C and Kremer, K},
doi = {10.1039/b919800h},
isbn = {1364-5498},
issn = {1359-6640},
journal = {Faraday Discussions},
pages = {9},
pmid = {20158020},
title = {{Multiscale simulation of soft matter systems}},
url = {http://xlink.rsc.org/?DOI=b919800h},
volume = {144},
year = {2010}
}



@article{Schlijper_1995,
abstract = {A novel method of investigating the link between molecular features of polymer molecules and the rheological properties of dilute polymer solutions has been investigated. It applies the dissipative particle dynamics (DPD) computer simulation technique, which introduces a lattice-gas automata time-stepping procedure into a molecular-dynamics scheme, to model bead-and-spring-type representations of polymer chains. Investigations of static and dynamic scaling relationships show that the scaling of radius of gyration and relaxation time with the number of beads are consistent with the predictions of the Rouse-Zimm model. Both hydrodynamic interaction and excluded volume emerge naturally from the DPD polymer model, indicating that a realistic description of the dynamics of a dilute polymer solution can be obtained within this framework, and that very efficient computer simulations are possible.},
author = {a.G. Schlijper and Hoogerbrugge, P.J. and Manke, C.W.},
doi = {10.1122/1.550713},
issn = {01486055},
journal = {Journal of Rheology},
number = {3},
pages = {567},
title = {{Computer simulation of dilute polymer solutions with the dissipative particle dynamics method}},
url = {https://www.engineeringvillage.com/share/document.url?mid=cpx\_4712679\&database=cpx},
volume = {39},
year = {1995}
}


@article{Spenley_EPL_2000,
abstract = {A nem technique, dissipative particle dynamics (DPD), appears promising as a means of studying the dynamical behaviour of polymers. Real polymers are known to obey a number of scaling laws, and a simulation method should reproduce these if it is to be relied on. The present work is a study of the properties of polymers, as simulated by DPD. Two cases are of interest: polymers in dilute solution, and polymers in the melt. The polymer in a good solvent. shows satisfactory agreement with scaling and Kirkwood theory, and the polymer melt is in excellent agreement with the predictions of Rouse theory},
author = {Spenley, N. a.},
doi = {10.1209/epl/i2000-00183-2},
isbn = {0295-5075},
issn = {0295-5075},
journal = {Europhysics Letters (EPL)},
number = {4},
pages = {534--540},
title = {{Scaling laws for polymers in dissipative particle dynamics}},
volume = {49},
year = {2000}
}


@article{Travis_JCP_2007,
abstract = {We introduce an improved method of parametrizing the Groot-Warren version of dissipative particle dynamics (DPD) by exploiting a correspondence between DPD and Scatchard-Hildebrand regular solution theory. The new parametrization scheme widens the realm of applicability of DPD by first removing the restriction of equal repulsive interactions between like beads, and second, by relating all conservative interactions between beads directly to cohesive energy densities. We establish the correspondence by deriving an expression for the Helmoltz free energy of mixing, obtaining a heat of mixing which is exactly the same form as that for a regular mixture (quadratic in the volume fraction) and an entropy of mixing which reduces to the ideal entropy of mixing for equal molar volumes. We equate the conservative interaction parameters in the DPD force law to the cohesive energy densities of the pure fluids, providing an alternative method of calculating the self-interaction parameters as well as a route to the cross interaction parameter. We validate the new parametrization by modeling the binary system SnI(4)SiCl(4), which displays liquid-liquid coexistence below an upper critical solution temperature around 140 degrees C. A series of DPD simulations were conducted at a set of temperatures ranging from 0 degrees C to above the experimental upper critical solution temperature using conservative parameters based on extrapolated experimental data. These simulations can be regarded as being equivalent to a quench from a high temperature to a lower one at constant volume. Our simulations recover the expected phase behavior ranging from solid-liquid coexistence to liquid-liquid coexistence and eventually leading to a homogeneous single phase system. The results yield a binodal curve in close agreement with the one predicted using regular solution theory, but, significantly, in closer agreement with actual solubility measurements.},
author = {Travis, Karl P. and Bankhead, Mark and Good, Kevin and Owens, Scott L.},
doi = {10.1063/1.2746325},
isbn = {0021-9606 (Print)$\backslash$r0021-9606 (Linking)},
issn = {00219606},
journal = {Journal of Chemical Physics},
number = {1},
pages = {0--12},
pmid = {17627339},
title = {{New parametrization method for dissipative particle dynamics}},
volume = {127},
year = {2007}
}


@article{Trofimov_JCP_2002,
abstract = {Dissipative particle dynamics (DPD) is a mesoscopic simulation method for studying hydrodynamic behavior of complex fluids. Ideally, a mesoscopic model should correctly represent the thermo- and hydrodynamic properties of a real system beyond certain length and time scales. Traditionally defined DPD quite successfully mimics hydrodynamics, but is not flexible enough to accurately describe the thermodynamics of a real system. The so-called “multibody” DPD (MDPD) is a pragmatic extension of the classical DPD that allows one to prescribe the thermodynamic behavior of a system with only a small performance impact. Here we present a practical improvement to the “multibody” DPD model and test it on a number of single-component examples. We also generalize MDPD to multicomponent systems, which are an important target of DPD studies. The improved model provides a correction for particle correlations in strongly nonideal systems that were neglected in the original MDPD model. The implications of the coarse-graining procedure on the MDPD are discussed.},
author = {Trofimov, S. Y. and Nies, E. L F and Michels, M. a J},
doi = {10.1063/1.1515774},
isbn = {9038618352},
issn = {00219606},
journal = {The Journal of Chemical Physics},
number = {20},
pages = {9383--9394},
title = {{Thermodynamic consistency in dissipative particle dynamics simulations of strongly nonideal liquids and liquid mixtures}},
volume = {117},
year = {2002}
}


@article{Warren_PRE_2003,
  title = {Vapor-liquid coexistence in many-body dissipative particle dynamics},
  author = {Warren, P. B.},
  journal = {Phys. Rev. E},
  volume = {68},
  issue = {6},
  pages = {066702},
  numpages = {8},
  year = {2003},
  month = {Dec},
  publisher = {American Physical Society},
  doi = {10.1103/PhysRevE.68.066702},
  url = {https://link.aps.org/doi/10.1103/PhysRevE.68.066702}
}


@article{Warren_PRE_2013,
  title = {No-go theorem in many-body dissipative particle dynamics},
  author = {Warren, Patrick B.},
  journal = {Phys. Rev. E},
  volume = {87},
  issue = {4},
  pages = {045303},
  numpages = {2},
  year = {2013},
  month = {Apr},
  publisher = {American Physical Society},
  doi = {10.1103/PhysRevE.87.045303},
  url = {https://link.aps.org/doi/10.1103/PhysRevE.87.045303}
}


@article{Wijmans_JCP_2001,
abstract = {We present Gibbs ensemble Monte Carlo simulations of monomer–solvent and polymer–solvent mixtures with soft interaction potentials, that are used in dissipative particle dynamics simulations. From the simulated phase behavior of the monomer–solvent mixtures one can derive an effective Flory–Huggins -parameter as a function of the particle interaction potential. We show that this -parameter agrees very well with the free energy difference between a monomer surrounded by solvent particles, and a solvent particle surrounded by solvent particles. We develop a new "identity change" Monte Carlo move to equilibrate the polymer–solvent mixtures. In this move a polymer chain from one box is exchanged with an equal number of solvent particles from the other box. At realistic densities this new move offers a large computational advantage over the convential insertion method for a polymer chain using a configurational bias Monte Carlo algorithm. The new algorithm is demonstrated for polymer–solvent mixtures with a chain length of up to 150 segments. Significant differences are found between the simulated polymer–solvent phase behavior and results predicted by mean-field theory. Finally, we fit a master–equation to the simulated binodal curves at different chain lengths. This function is used to make a quantitative comparison between the simulations and experimental data for the phase equilibrium of the polystyrene–methylcyclohexane system.},
author = {Wijmans, C. M. and Smit, B. and Groot, R. D.},
doi = {10.1063/1.1362298},
isbn = {0021-9606},
issn = {00219606},
journal = {Journal of Chemical Physics},
number = {17},
pages = {7644--7654},
pmid = {168168100040},
title = {{Phase behavior of monomeric mixtures and polymer solutions with soft interaction potentials}},
volume = {114},
year = {2001}
}


@article{DLMS,
author = {Michael A. Seaton and Richard L. Anderson and Sebastian Metz and William Smith},
title = {DL\_MESO: highly scalable mesoscale simulations},
journal = {Molecular Simulation},
volume = {39},
number = {10},
pages = {796-821},
year = {2013},
doi = {10.1080/08927022.2013.772297},
URL = {http://dx.doi.org/10.1080/08927022.2013.772297},
eprint = {http://dx.doi.org/10.1080/08927022.2013.772297},
abstract = { DL\_MESO is a parallel mesoscale simulation package capable of dissipative particle dynamics and the lattice Boltzmann equation method. It has been developed at Daresbury Laboratory for the United Kingdom Collaborative Computational Project known as CCP5. Capable of addressing industrially relevant tasks, but written to support academic research, it has a wide range of applications and scales to thousands of processors on high-performance computing platforms yet runs efficiently on smaller commodity clusters and single processor personal computers. This article serves as a guide to a variety of users, describing the functionality, performance and structure of this simulation package. Representative examples highlighting the capabilities of DL\_MESO are given for each of the two methodologies available. Future directions for the package are discussed towards the end of the article. }
}


@article{Fuchslin_JCP_2009,
author = {Rudolf M. F\"uchslin and Harold Fellermann and Anders Eriksson and Hans-Joachim Ziock},
title = {Coarse graining and scaling in dissipative particle dynamics},
journal = {The Journal of Chemical Physics},
volume = {130},
number = {21},
pages = {214102},
year = {2009},
doi = {10.1063/1.3143976},
URL = {http://dx.doi.org/10.1063/1.3143976},
eprint = {http://dx.doi.org/10.1063/1.3143976}
}


@article{Kendrick_JACS_2010,
author = {Kendrick, Ian and Kumari, Dunesh and Yakaboski, Adam and Dimakis, Nicholas and Smotkin, Eugene S.},
title = {Elucidating the Ionomer-Electrified Metal Interface},
journal = {Journal of the American Chemical Society},
volume = {132},
number = {49},
pages = {17611-17616},
year = {2010},
doi = {10.1021/ja1081487},
note ={PMID: 21087013},
URL = {http://dx.doi.org/10.1021/ja1081487},
eprint = {http://dx.doi.org/10.1021/ja1081487}
}


@article{Kim_MA_2013,
author = {Kim, Sangcheol and Dura, Joseph A. and Page, Kirt A. and Rowe, Brandon W. and Yager, Kevin G. and Lee, Hae-Jeong and Soles, Christopher L.},
title = {Surface-Induced Nanostructure and Water Transport of Thin Proton-Conducting Polymer Films},
journal = {Macromolecules},
volume = {46},
number = {14},
pages = {5630-5637},
year = {2013},
doi = {10.1021/ma400750f},
URL = {http://dx.doi.org/10.1021/ma400750f},
eprint = {http://dx.doi.org/10.1021/ma400750f}
}


@article{Modestino_MA_2013,
author = {Modestino, Miguel A. and Paul, Devproshad K. and Dishari, Shudipto and Petrina, Stephanie A. and Allen, Frances I. and Hickner, Michael A. and Karan, Kunal and Segalman, Rachel A. and Weber, Adam Z.},
title = {Self-Assembly and Transport Limitations in Confined Nafion Films},
journal = {Macromolecules},
volume = {46},
number = {3},
pages = {867-873},
year = {2013},
doi = {10.1021/ma301999a},
URL = {http://dx.doi.org/10.1021/ma301999a},
eprint = {http://dx.doi.org/10.1021/ma301999a}
}


@article{Kusoglu_ChemRev_2017,
author = {Kusoglu, Ahmet and Weber, Adam Z.},
title = {New Insights into Perfluorinated Sulfonic-Acid Ionomers},
journal = {Chemical Reviews},
volume = {117},
number = {3},
pages = {987-1104},
year = {2017},
doi = {10.1021/acs.chemrev.6b00159},
note ={PMID: 28112903},
URL = {http://dx.doi.org/10.1021/acs.chemrev.6b00159},
eprint = {http://dx.doi.org/10.1021/acs.chemrev.6b00159}
}


@article{Espanol_JCP_2017,
author = {Pep Espa{\~n}ol and Patrick B. Warren},
title = {Perspective: Dissipative particle dynamics},
journal = {The Journal of Chemical Physics},
volume = {146},
number = {15},
pages = {150901},
year = {2017},
doi = {10.1063/1.4979514},
URL = {http://dx.doi.org/10.1063/1.4979514},
}


@article{Atashafrooz_JCED_2016,
author = {, Maryam and Mehdipour, Nargess},
title = {Many-Body Dissipative Particle Dynamics Simulation of Liquid–Vapor Coexisting Curve in Sodium},
journal = {Journal of Chemical \& Engineering Data},
volume = {61},
number = {10},
pages = {3659-3664},
year = {2016},
doi = {10.1021/acs.jced.6b00586},

URL = { 
        http://dx.doi.org/10.1021/acs.jced.6b00586
    
},
eprint = { 
        http://dx.doi.org/10.1021/acs.jced.6b00586
    
}

}


@article{Jamali_JCP_2015,
author = {Safa Jamali and Arman Boromand and Shaghayegh Khani and Jacob Wagner and Mikio Yamanoi and Joao Maia},
title = {Generalized mapping of multi-body dissipative particle dynamics onto fluid compressibility and the Flory-Huggins theory},
journal = {The Journal of Chemical Physics},
volume = {142},
number = {16},
pages = {164902},
year = {2015},
doi = {10.1063/1.4919303},

URL = { 
        https://doi.org/10.1063/1.4919303
    
},
eprint = { 
        https://doi.org/10.1063/1.4919303
    
}

}


@article{Kadoya_PRE_2017,
  title = {Size dependence of static polymer droplet behavior from many-body dissipative particle dynamics simulation},
  author = {Kadoya, Naoki and Arai, Noriyoshi},
  journal = {Phys. Rev. E},
  volume = {95},
  issue = {4},
  pages = {043109},
  numpages = {10},
  year = {2017},
  month = {Apr},
  publisher = {American Physical Society},
  doi = {10.1103/PhysRevE.95.043109},
  url = {https://link.aps.org/doi/10.1103/PhysRevE.95.043109}
}

@article{Yamamoto_JCP_2002,
author = {Satoru Yamamoto and Yutaka Maruyama and Shi-aki Hyodo},
title = {Dissipative particle dynamics study of spontaneous vesicle formation of amphiphilic molecules},
journal = {The Journal of Chemical Physics},
volume = {116},
number = {13},
pages = {5842-5849},
year = {2002},
doi = {10.1063/1.1456031},

URL = { 
        https://doi.org/10.1063/1.1456031
    
},
eprint = { 
        https://doi.org/10.1063/1.1456031
    
}

}


@Article{Yong_Polymers_2016,
AUTHOR = {Yong, Xin},
TITLE = {Hydrodynamic Interactions and Entanglements of Polymer Solutions in Many-Body Dissipative Particle Dynamics},
JOURNAL = {Polymers},
VOLUME = {8},
YEAR = {2016},
NUMBER = {12},
URL = {http://www.mdpi.com/2073-4360/8/12/426},
ISSN = {2073-4360},
ABSTRACT = {Using many-body dissipative particle dynamics (MDPD), polymer solutions with concentrations spanning dilute and semidilute regimes are modeled. The parameterization of MDPD interactions for systems with liquid–vapor coexistence is established by mapping to the mean-field Flory–Huggins theory. The characterization of static and dynamic properties of polymer chains is focused on the effects of hydrodynamic interactions and entanglements. The coil–globule transition of polymer chains in dilute solutions is probed by varying solvent quality and measuring the radius of gyration and end-to-end distance. Both static and dynamic scaling relations for polymer chains in poor, theta, and good solvents are in good agreement with the Zimm theory with hydrodynamic interactions considered. Semidilute solutions with polymer volume fractions up to 0.7 exhibit the screening of excluded volume interactions and subsequent shrinking of polymer coils. Furthermore, entanglements become dominant in the semidilute solutions, which inhibit diffusion and relaxation of chains. Quantitative analysis of topology violation confirms that entanglements are correctly captured in the MDPD simulations.},
DOI = {10.3390/polym8120426}
}



@article{Vega_JCP_2007,
author = {C. Vega and E. de Miguel},
title = {Surface tension of the most popular models of water by using the test-area simulation method},
journal = {The Journal of Chemical Physics},
volume = {126},
number = {15},
pages = {154707},
year = {2007},
doi = {10.1063/1.2715577},

URL = { 
        https://doi.org/10.1063/1.2715577
    
},
eprint = { 
        https://doi.org/10.1063/1.2715577
    
}

}


@article{Vanya_JCP_2017,
author = {P. Vanya and J. Sharman and J. A. Elliott},
title = {Mesoscale simulations of confined Nafion thin films},
journal = {The Journal of Chemical Physics},
volume = {147},
number = {21},
pages = {214904},
year = {2017},
doi = {10.1063/1.4996695},

URL = { 
        https://doi.org/10.1063/1.4996695
    
},
eprint = { 
        https://doi.org/10.1063/1.4996695
    
}

}


@article{Maiti_JCP_2004,
author = {Amitesh Maiti and Simon McGrother},
title = {Bead–bead interaction parameters in dissipative particle dynamics: Relation to bead-size, solubility parameter, and surface tension},
journal = {The Journal of Chemical Physics},
volume = {120},
number = {3},
pages = {1594-1601},
year = {2004},
doi = {10.1063/1.1630294},

URL = { 
        https://doi.org/10.1063/1.1630294
    
},
eprint = { 
        https://doi.org/10.1063/1.1630294
    
}

}


@article{Izvekov_JCP_2014,
author = {Sergei Izvekov and Betsy M. Rice},
title = {Multi-scale coarse-graining of non-conservative interactions in molecular liquids},
journal = {The Journal of Chemical Physics},
volume = {140},
number = {10},
pages = {104104},
year = {2014},
doi = {10.1063/1.4866142},
URL = {https://doi.org/10.1063/1.4866142}
}


@article{LAMMPS,
title = "Fast Parallel Algorithms for Short-Range Molecular Dynamics",
journal = "Journal of Computational Physics",
volume = "117",
number = "1",
pages = "1 - 19",
year = "1995",
issn = "0021-9991",
doi = "https://doi.org/10.1006/jcph.1995.1039",
url = "http://www.sciencedirect.com/science/article/pii/S002199918571039X",
author = "Steve Plimpton"
}


@book{Trofimov_PhD,
author = {Trofimov, S.},
title = {Thermodynamic consistency in dissipative particle dynamics},
publisher = {PhD thesis, Technische Universiteit Eindhoven},
year = {2003}
}


@article{Vanya_PRE_2018,
  title = {Liquid-phase parametrization and solidification in many-body dissipative particle dynamics},
  author = {Vanya, Peter and Crout, Phillip and Sharman, Jonathan and Elliott, James A.},
  journal = {Phys. Rev. E},
  volume = {98},
  issue = {3},
  pages = {033310},
  numpages = {9},
  year = {2018},
  month = {Sep},
  publisher = {American Physical Society},
  doi = {10.1103/PhysRevE.98.033310},
  url = {https://link.aps.org/doi/10.1103/PhysRevE.98.033310}
}


@article{Merabia_JCP_2007,
author = {Merabia,Samy  and Pagonabarraga,Ignacio },
title = {Density dependent potentials: Structure and thermodynamics},
journal = {The Journal of Chemical Physics},
volume = {127},
number = {5},
pages = {054903},
year = {2007},
doi = {10.1063/1.2751496},
URL = {https://doi.org/10.1063/1.2751496},
}


@article{Vanya_JCP_2019,
author = {Vanya,Peter  and Sharman,Jonathan  and Elliott,James A. },
title = {Invariance of experimental observables with respect to coarse-graining in standard and many-body dissipative particle dynamics},
journal = {The Journal of Chemical Physics},
volume = {150},
number = {6},
pages = {064101},
year = {2019},
doi = {10.1063/1.5046851},
URL = {https://doi.org/10.1063/1.5046851},
}


@article{Louis_JPCM_2002,
	doi = {10.1088/0953-8984/14/40/311},
	url = {https://doi.org/10.1088
	year = 2002,
	month = {sep},
	publisher = {{IOP} Publishing},
	volume = {14},
	number = {40},
	pages = {9187--9206},
	author = {A A Louis},
	title = {Beware of density dependent pair potentials},
	journal = {Journal of Physics: Condensed Matter},
}


@article{Sanyal_JPCB_2018,
author = {Sanyal, Tanmoy and Shell, M. Scott},
title = {Transferable Coarse-Grained Models of Liquid–Liquid Equilibrium Using Local Density Potentials Optimized with the Relative Entropy},
journal = {The Journal of Physical Chemistry B},
volume = {122},
number = {21},
pages = {5678-5693},
year = {2018},
doi = {10.1021/acs.jpcb.7b12446},
note ={PMID: 29466859},
URL = {https://doi.org/10.1021/acs.jpcb.7b12446}
}


@article{Sanyal_JCP_2016,
author = {Sanyal,Tanmoy  and Shell,M. Scott },
title = {Coarse-grained models using local-density potentials optimized with the relative entropy: Application to implicit solvation},
journal = {The Journal of Chemical Physics},
volume = {145},
number = {3},
pages = {034109},
year = {2016},
doi = {10.1063/1.4958629},
URL = {https://doi.org/10.1063/1.4958629}
}


@article{Allen_JCP_2008,
author = {Allen,Erik C.  and Rutledge,Gregory C. },
title = {A novel algorithm for creating coarse-grained, density dependent implicit solvent models},
journal = {The Journal of Chemical Physics},
volume = {128},
number = {15},
pages = {154115},
year = {2008},
doi = {10.1063/1.2899729},
URL = {https://doi.org/10.1063/1.2899729}
}


@article{Allen_JCP_2009,
author = {Allen,Erik C.  and Rutledge,Gregory C. },
title = {Evaluating the transferability of coarse-grained, density-dependent implicit solvent models to mixtures and chains},
journal = {The Journal of Chemical Physics},
volume = {130},
number = {3},
pages = {034904},
year = {2009},
doi = {10.1063/1.3055594},
URL = {https://doi.org/10.1063/1.3055594}
}


@article{Daw_PRL_1983,
  title = {Semiempirical, Quantum Mechanical Calculation of Hydrogen Embrittlement in Metals},
  author = {Daw, Murray S. and Baskes, M. I.},
  journal = {Phys. Rev. Lett.},
  volume = {50},
  issue = {17},
  pages = {1285--1288},
  numpages = {0},
  year = {1983},
  month = {Apr},
  publisher = {American Physical Society},
  doi = {10.1103/PhysRevLett.50.1285},
  url = {https://link.aps.org/doi/10.1103/PhysRevLett.50.1285}
}


@article{Daw_PRB_1984,
  title = {Embedded-atom method: Derivation and application to impurities, surfaces, and other defects in metals},
  author = {Daw, Murray S. and Baskes, M. I.},
  journal = {Phys. Rev. B},
  volume = {29},
  issue = {12},
  pages = {6443--6453},
  numpages = {0},
  year = {1984},
  month = {Jun},
  publisher = {American Physical Society},
  doi = {10.1103/PhysRevB.29.6443},
  url = {https://link.aps.org/doi/10.1103/PhysRevB.29.6443}
}


@article{Johnson_PRB_1989,
  title = {Alloy models with the embedded-atom method},
  author = {Johnson, R. A.},
  journal = {Phys. Rev. B},
  volume = {39},
  issue = {17},
  pages = {12554--12559},
  numpages = {0},
  year = {1989},
  month = {Jun},
  publisher = {American Physical Society},
  doi = {10.1103/PhysRevB.39.12554},
  url = {https://link.aps.org/doi/10.1103/PhysRevB.39.12554}
}


@article{Finnis_PMA_1984,
author = { M. W.   Finnis  and  J. E.   Sinclair },
title = {A simple empirical N-body potential for transition metals},
journal = {Philosophical Magazine A},
volume = {50},
number = {1},
pages = {45-55},
year  = {1984},
publisher = {Taylor & Francis},
doi = {10.1080/01418618408244210},
URL = {https://doi.org/10.1080/01418618408244210}
}


@article{Sutton_PML_1990,
author = { A. P.   Sutton  and  J.   Chen },
title = {Long-range Finnis–Sinclair potentials},
journal = {Philosophical Magazine Letters},
volume = {61},
number = {3},
pages = {139-146},
year  = {1990},
publisher = {Taylor & Francis},
doi = {10.1080/09500839008206493},
URL = {https://doi.org/10.1080/09500839008206493}
}



@article{RafiiTabar_PML_1991,
author = { H.   Rafii-Tabar  and  A. P.   Sulton },
title = {Long-range Finnis-Sinclair potentials for f.c.c. metallic alloys},
journal = {Philosophical Magazine Letters},
volume = {63},
number = {4},
pages = {217-224},
year  = {1991},
publisher = {Taylor & Francis},
doi = {10.1080/09500839108205994},
URL = {https://doi.org/10.1080/09500839108205994}
}


@article{Warren_PRL_2001,
  title = {Hydrodynamic Bubble Coarsening in Off-Critical Vapor-Liquid Phase Separation},
  author = {Warren, Patrick B.},
  journal = {Phys. Rev. Lett.},
  volume = {87},
  issue = {22},
  pages = {225702},
  numpages = {4},
  year = {2001},
  month = {Nov},
  publisher = {American Physical Society},
  doi = {10.1103/PhysRevLett.87.225702},
  url = {https://link.aps.org/doi/10.1103/PhysRevLett.87.225702}
}




@article{Espanol_PRE_2003,
  title = {Smoothed dissipative particle dynamics},
  author = {Espa\~nol, Pep and Revenga, Mariano},
  journal = {Phys. Rev. E},
  volume = {67},
  issue = {2},
  pages = {026705},
  numpages = {12},
  year = {2003},
  month = {Feb},
  publisher = {American Physical Society},
  doi = {10.1103/PhysRevE.67.026705},
  url = {https://link.aps.org/doi/10.1103/PhysRevE.67.026705}
}


@article{Litvinov_PRE_2008,
  title = {Smoothed dissipative particle dynamics model for polymer molecules in suspension},
  author = {Litvinov, Sergey and Ellero, Marco and Hu, Xiangyu and Adams, Nikolaus A.},
  journal = {Phys. Rev. E},
  volume = {77},
  issue = {6},
  pages = {066703},
  numpages = {12},
  year = {2008},
  month = {Jun},
  publisher = {American Physical Society},
  doi = {10.1103/PhysRevE.77.066703},
  url = {https://link.aps.org/doi/10.1103/PhysRevE.77.066703}
}


@article{Thieulot_PRE_2005a,
  title = {Smoothed particle hydrodynamics model for phase separating fluid mixtures. I. General equations},
  author = {Thieulot, Cedric and Janssen, L. P. B. M. and Espa\~nol, Pep},
  journal = {Phys. Rev. E},
  volume = {72},
  issue = {1},
  pages = {016713},
  numpages = {15},
  year = {2005},
  month = {Jul},
  publisher = {American Physical Society},
  doi = {10.1103/PhysRevE.72.016713},
  url = {https://link.aps.org/doi/10.1103/PhysRevE.72.016713}
}


@article{Thieulot_PRE_2005b,
  title = {Smoothed particle hydrodynamics model for phase separating fluid mixtures. II. Diffusion in a binary mixture},
  author = {Thieulot, Cedric and Janssen, L. P. B. M. and Espa\~nol, Pep},
  journal = {Phys. Rev. E},
  volume = {72},
  issue = {1},
  pages = {016714},
  numpages = {12},
  year = {2005},
  month = {Jul},
  publisher = {American Physical Society},
  doi = {10.1103/PhysRevE.72.016714},
  url = {https://link.aps.org/doi/10.1103/PhysRevE.72.016714}
}


@article{Nugent_PRE_2000,
  title = {Liquid drops and surface tension with smoothed particle applied mechanics},
  author = {Nugent, S. and Posch, H. A.},
  journal = {Phys. Rev. E},
  volume = {62},
  issue = {4},
  pages = {4968--4975},
  numpages = {0},
  year = {2000},
  month = {Oct},
  publisher = {American Physical Society},
  doi = {10.1103/PhysRevE.62.4968},
  url = {https://link.aps.org/doi/10.1103/PhysRevE.62.4968}
}



@article{Espanol_JCP_2003,
author = {Español,Pep  and Thieulot,Cedric },
title = {Microscopic derivation of hydrodynamic equations for phase-separating fluid mixtures},
journal = {The Journal of Chemical Physics},
volume = {118},
number = {20},
pages = {9109-9127},
year = {2003},
doi = {10.1063/1.1568333},
URL = {https://doi.org/10.1063/1.1568333}
}


@article{Merabia_PRL_2008,
  title = {Dewetting of a Stratified Two-Component Liquid Film on a Solid Substrate},
  author = {Merabia, Samy and Avalos, Josep Bonet},
  journal = {Phys. Rev. Lett.},
  volume = {101},
  issue = {20},
  pages = {208304},
  numpages = {4},
  year = {2008},
  month = {Nov},
  publisher = {American Physical Society},
  doi = {10.1103/PhysRevLett.101.208304},
  url = {https://link.aps.org/doi/10.1103/PhysRevLett.101.208304}
}



@article{Merabia_EPJE_2006,
title = {A mesoscopic model for (de)wetting},
author = {Merabia, S. and Pagonabarraga, I.},
journal = {The European Physical Journal E},
volume = {20},
issue = {2},
pages = {209-214},
year = {2006},
url = {https://doi.org/10.1140/epje/i2005-10128-1},
doi = {10.1140/epje/i2005-10128-1}
}
	

@article{Merabia_JNNFM_2008,
title = "Modelling capillary phenomena at a mesoscale: From simple to complex fluids",
journal = "Journal of Non-Newtonian Fluid Mechanics",
volume = "154",
number = "1",
pages = "13 - 21",
year = "2008",
issn = "0377-0257",
doi = "https://doi.org/10.1016/j.jnnfm.2008.01.009",
url = "http://www.sciencedirect.com/science/article/pii/S0377025708000116",
author = "S. Merabia and J. Bonet-Avalos and I. Pagonabarraga",
keywords = "Capillary phenomena, Mesoscopic drop model, Thin films, Binary mixtures",
abstract = "A mesoscopic model for studying capillary phenomena is introduced. The fluid is represented by particles interacting through soft forces that allow condensation. A model for a solid wall is also presented whose affinity for the liquid can be tuned from hydrophilic to superhydrophobic. Regarding the dynamics, the validity of the model was assessed studying the classical drop spreading on a wetting substrate where good agreement was found with the scaling predicted theoretically. We show also how to extend the proposed model to deal with symmetrical binary mixtures. This model opens the way to model capillary phenomena involving complex fluids."
}


@Article{Li_Nanoscale_2012,
author ="Li, Yinfeng and Li, Xuejin and Li, Zhonghua and Gao, Huajian",
title  ="Surface-structure-regulated penetration of nanoparticles across a cell membrane",
journal  ="Nanoscale",
year  ="2012",
volume  ="4",
issue  ="12",
pages  ="3768-3775",
publisher  ="The Royal Society of Chemistry",
doi  ="10.1039/C2NR30379E",
url  ="http://dx.doi.org/10.1039/C2NR30379E",
}


@article{Li_MA_2009,
author = {Li, Xuejin and Pivkin, Igor V. and Liang, Haojun and Karniadakis, George Em},
title = {Shape Transformations of Membrane Vesicles from Amphiphilic Triblock Copolymers: A Dissipative Particle Dynamics Simulation Study},
journal = {Macromolecules},
volume = {42},
number = {8},
pages = {3195-3200},
year = {2009},
doi = {10.1021/ma9000918},
URL = {https://doi.org/10.1021/ma9000918}
}


@article {Peng_PNAS_2013,
	author = {Peng, Zhangli and Li, Xuejin and Pivkin, Igor V. and Dao, Ming and Karniadakis, George E. and Suresh, Subra},
	title = {Lipid bilayer and cytoskeletal interactions in a red blood cell},
	volume = {110},
	number = {33},
	pages = {13356--13361},
	year = {2013},
	doi = {10.1073/pnas.1311827110},
	publisher = {National Academy of Sciences},
	issn = {0027-8424},
	URL = {https://www.pnas.org/content/110/33/13356},
	eprint = {https://www.pnas.org/content/110/33/13356.full.pdf},
	journal = {Proceedings of the National Academy of Sciences}
}


@article{Li_PTRSA_2014,
author = {Li, Xuejin  and Peng, Zhangli  and Lei, Huan  and Dao, Ming  and Karniadakis, George Em },
title = {Probing red blood cell mechanics, rheology and dynamics with a two-component multi-scale model},
journal = {Philosophical Transactions of the Royal Society A: Mathematical, Physical and Engineering Sciences},
volume = {372},
number = {2021},
pages = {20130389},
year = {2014},
doi = {10.1098/rsta.2013.0389},
URL = {https://royalsocietypublishing.org/doi/abs/10.1098/rsta.2013.0389},
}


@article{Chang_PLOS_2016,
    author = {Chang, Hung-Yu AND Li, Xuejin AND Li, He AND Karniadakis, George Em},
    journal = {PLOS Computational Biology},
    publisher = {Public Library of Science},
    title = {MD/DPD Multiscale Framework for Predicting Morphology and Stresses of Red Blood Cells in Health and Disease},
    year = {2016},
    month = {10},
    volume = {12},
    url = {https://doi.org/10.1371/journal.pcbi.1005173},
    pages = {1-22},
    number = {10},
    doi = {10.1371/journal.pcbi.1005173}
}

\end{document}